\documentclass[a4paper,11pt]{article}

\pdfoutput=1 

\usepackage{jheppub} 

\usepackage[T1]{fontenc} 
\usepackage{subcaption}
\usepackage{verbatim}
\usepackage{mathrsfs}
\usepackage{physics}
\usepackage{amsthm}
\usepackage{mathtools}
\usepackage{enumerate}
\usepackage{graphicx} 

\usepackage{amsmath,amssymb}
\usepackage{xcolor}
\usepackage{tikz}
\usetikzlibrary{calc,fadings}

\definecolor{ForestGreen}{RGB}{34,139,34}
\definecolor{ArFill}{RGB}{142,108,207}
\definecolor{AsFill}{RGB}{244,177,131}
\definecolor{AspFill}{RGB}{143,183,255}

\definecolor{ArLine}{RGB}{111,66,193}
\definecolor{AsLine}{RGB}{230,126,34}
\definecolor{AspLine}{RGB}{58,134,255}

\tikzfading[
  name=fadeRight,
  left color=transparent!0,
  right color=transparent!100
]

\tikzfading[
  name=fadeLeft,
  left color=transparent!100,
  right color=transparent!0
]
\usepackage{amsmath,amssymb}
\usepackage{dsfont}

\renewcommand{\tilde}{\widetilde}
\renewcommand{\bar}{\overline}

\renewcommand{\Im}{\text{Im}\,}

\newcommand{\im}{\operatorname{im}}

\DeclareMathOperator{\supp}{supp}

\newtheoremstyle{breakthm}%
{}{}%
{\itshape}{}%
{\bfseries}{}
{\newline}{}

\theoremstyle{breakthm}
\newtheorem{theorem}{Theorem}

\newtheorem{prop}[theorem]{Proposition}

\newtheoremstyle{break}%
{\topsep}{\topsep}%
{}{}%
{\bfseries}{}
{\newline}{}

\theoremstyle{break}

\numberwithin{theorem}{section}

\title{Modular conjugations for thermal states and  second worlds}

\author[a,b]{Alan Garbarz,}
\author[c,d]{Gabriel Palau,}
\author[a,b]{and Guillem P\'erez-Nadal}

\affiliation[a]{CONICET - Universidad de Buenos Aires, Instituto de Física de Buenos Aires (IFIBA).\\ Buenos Aires, Argentina
}
\affiliation[b]{Universidad de Buenos Aires, Facultad de Ciencias Exactas y Naturales, Departamento de Física.\\ Buenos Aires, Argentina.
}
\affiliation[c]{Universidad Argentina de la Empresa (UADE), Lima 717, Buenos Aires, Argentina.}
\affiliation[d]{FIUBA, Paseo Colón 850, Buenos Aires, Argentina.}

\emailAdd{alan@df.uba.ar}
\emailAdd{gbrlpl8@gmail.com}
\emailAdd{guillem@df.uba.ar}

\abstract{The modular conjugation $J$ maps an algebra of observables onto its commutant. For the vacuum on a Rindler wedge it is a spacetime reflection, but for thermal states in two-dimensional CFTs the known expressions involve chiral fields at the complex points $x\pm i\beta/2$, interpreted as belonging to a ``second world''. Here we compute $J$ for the geometric thermal state in explicit cyclic purifications, using either the vacuum vector, through the conformal map to a Rindler wedge, or a vector in the natural cone of a local algebra. The modular images are then fields at real points, and complex coordinates arise only when operators of the commutant outside the image of the thermal representation are written as fields on the original spacetime. For the chiral $U(1)$ current we also purify by doubling the Fock space, obtain explicit formulas for $J$, show that this purification coincides with Sorce's canonical purification, and treat all primary thermal states.}

\begin{document}
\maketitle

\section{Introduction}

The modular conjugation is an antiunitary operator associated with an algebra and a cyclic and separating vector. Its main property is that it maps the algebra to its commutant. In quantum field theory, where algebras are assigned to spacetime regions, this raises the question of where the transformed operators are localized. For the vacuum there are important examples in which the answer is geometric. For a thermal state the situation is more involved, since the commutant can contain operators which do not correspond to fields in the original representation of spacetime.

The best-known vacuum result is the Bisognano--Wichmann theorem, which identifies the modular evolution $\Delta^{it}$ of a Rindler wedge with Lorentz boosts and the modular conjugation $J$ with a spacetime reflection, together with the appropriate action on the fields \cite{Bisognano:1975ih,Bisognano:1976za}. The modular operators of a double cone were obtained by Hislop and Longo for the free massless scalar field \cite{HislopLongo1982}, and geometric modular action in conformal field theory was studied more generally by Brunetti, Guido and Longo \cite{BrunettiGuidoLongo1993}. At finite temperature, Borchers and Yngvason found geometric modular flows for wedges and related regions in certain two-dimensional theories \cite{BorchersYngvason1999}. Schroer and Wiesbrock showed how a thermal chiral system and its commutant can be realized within a ground-state theory on a larger spacetime, with the commutant assigned to a region beyond a horizon \cite{SchroerWiesbrock2000}. This is the seed of the terminology ``second world'', introduced by Borchers in \cite{Borchers:1999thermal}, where localization on a larger space-time than the original takes place.

More recently, Mintchev and Tonni studied the geometric action of modular conjugation for an interval in a two-dimensional CFT, both in the vacuum and in a thermal state \cite{Mintchev:2022fcp}, by analytical continuation of the modular evolution to $t=-i\beta/2$. In the thermal case, part of the interval is mapped to its complement on the original line, while another part acquires an imaginary displacement $\pm i\beta/2$. They interpreted this latter contribution in terms of a second world and related the inversion on a spatial slice to geodesic bit threads in the BTZ black brane geometry. In particular, the threads associated with that part of the interval reach the horizon. Their work gives a concrete geometric description of the additional region required by the thermal commutant that we will find again. 

Explicit field transformations were obtained by Abate, Blanco, Koifman and P\'erez-Nadal for the vacuum of a massless Dirac field in multicomponent regions \cite{AbateEtAl2023}. The modular conjugation mixes fields at different points, and the result was used to analyze Haag duality. Abate and Koifman subsequently studied the thermal chiral fermion on a circle \cite{AbateKoifman2023}. Their expression contains a sum of contributions from the spatial complement and an additional term written using a field at a complex coordinate $x+i\beta/2$. It also exhibits nonlocality for a single interval. The additional term is interpreted as belonging, after the fermionic twist, to the commutant of the global thermal algebra. This interpretation also uses the aforementioned  analytical continuation of $\Delta^{it}$ in the space of local operators.

These results motivate a closer look at the meaning of expressions such as $\psi(x\pm i\beta/2)$. The KMS condition gives analyticity of correlation functions, and modular theory gives corresponding vector identities. For a bounded operator {$a$} in an algebra $\mathcal A$, with cyclic and separating vector $|\Psi\rangle$, one has
\begin{equation}\label{intro:Tomita-vector}
    (JaJ)|\Psi\rangle=\Delta^{1/2}a^\dagger|\Psi\rangle.
\end{equation}
If one already has an operator $b\in\mathcal A'$ with this action on $|\Psi\rangle$, then $b=JaJ$, because $|\Psi\rangle$ is separating for $\mathcal A'$. However, analytically continuing the evolution of $a^\dagger$ does not by itself produce such an operator in the commutant. Agreement on the cyclic vector alone, as in \cite{AbateKoifman2023}, does not establish an operator identity, as we next discuss\footnote{We thank N. Abate and M. Koifman for {correspondence} about this.}.

This distinction can already be seen for an invertible density matrix in finite dimension. Consider its purification by $\rho^{1/2}$ in the space of Hilbert--Schmidt operators, with $\pi(a)s=as$. The modular conjugation is $Js=s^\dagger$, so
\begin{equation}\label{intro:HS-actions}
    J\pi(a)Js=sa^\dagger,\qquad
    \pi\!\left(\rho^{1/2}\,a^\dagger\rho^{-1/2}\right)s
    =\rho^{1/2}\,a^\dagger\rho^{-1/2}s.
\end{equation}
For $\rho=Z^{-1}e^{-\beta H}$, the second operator is precisely $\pi(\alpha_{i\beta/2}(a^\dagger))$, where $\alpha_t(a)=e^{itH}ae^{-itH}$. Equivalently, it is $\Delta^{it}\pi(a^\dagger)\Delta^{-it}$ continued to $t=-i/2$, since the modular group of a thermal state is the time translation by $-\beta t$. Both operators send $\rho^{1/2}$ to $\rho^{1/2}\,a^\dagger$, but their actions on a general $s$ differ. The first is right multiplication by $a^\dagger$ and belongs to the commutant. The second is a left multiplication and, unless $a$ is scalar, does not. Thus the issue is present even when analytic continuation exists for every operator and there are no domain subtleties.

For the thermal fermion, the identification of the complex-coordinate field with the purifying copy is invoked in passing from a vector identity to an operator identity in \cite{AbateKoifman2023}. One must either construct the proposed commutant field independently and check its action on the cyclic vector, or show that a specified operator continuation has the required commutation properties, including the fermionic twist. This leaves a question about the operator meaning of the complex-coordinate notation, which is the starting motivation here. 

The main goal of this note is to compute the modular conjugation in explicit cyclic purifications and, as a byproduct, use them to clarify this point. We consider the geometric KMS state of a two-dimensional CFT, constructed from the vacuum by a conformal transformation \cite{Camassa:2011te}. In a chiral component, the map $\sigma(x)=e^{2\pi x/\beta}$ identifies the thermal line with the positive vacuum half-line. We first compute $J$ in this vacuum realization, where the transformed fields have real coordinates. Complex coordinates appear when one tries to undo $\sigma$ on a reflected point outside its image. We then compare this construction with purifications in natural positive cones. Araki's results imply that the faithful vector representatives in a fixed natural cone share the same modular conjugation \cite{Araki:1972zza}. Using the cone of the local algebra, or that of a larger algebra, gives different realizations of the same local thermal state. We describe the corresponding actions of $J$ and where their images are localized.

Finally, we study the chiral $U(1)$ current, namely the derivative of a free massless chiral scalar. This model allows us to make the preceding discussion explicit at the one-particle level. Following the doubling construction for quasi-free states \cite{ArakiShiraishi1971}, we compute the modular conjugation on a doubled real space and obtain its action in the Fock representation. Localization enters through the standard subspaces generated by $f\oplus0$ with $f$ supported in the chosen interval. The formulas show directly that the modular images lie in the appropriate symplectic complements, hence that the corresponding Weyl operators lie in the commutants. For a half-line, they distinguish the functions whose images remain on the complementary thermal half-line from those whose images belong to the global commutant. This provides an explicit description in terms of smeared fields and Hilbert-space operators. It also complements the computation of thermal modular evolution for a bounded interval in \cite{GarbarzPalau2023}. Using the classification in \cite{Camassa:2011wk}, we further obtain local modular conjugations for all primary locally normal KMS states of the current in the representations constructed here.

{Our constructions are also related to the canonical purification of algebraic states recently introduced by Sorce \cite{sorce2025continuumcanonicalpurifications}, in which a state on an algebra $\mathcal{A}$ is purified on the doubled algebra $\mathcal{A}\otimes\mathcal{A}^{\rm op}$, and the opposite algebra is represented in the commutant by means of the modular conjugation. Each of the purifications considered here is a concrete realization of it in which the localization of the second copy can be read off. For the current we show explicitly that the doubled Fock purification coincides with the canonical purification of the Gaussian thermal state, in the form given in \cite{Caminiti:2026excitability}, and that the correlations between the two copies are given by the continuation of the thermal two-point function to the middle of the KMS strip.}

The paper is organized as follows. Section~\ref{sec:setup} fixes the notation, collects the facts about cyclic purifications and modular conjugation that we need, and outlines our strategy{, including a first explanation of where second worlds come from}. Section~\ref{2dCFTs} treats the geometric KMS state in the vacuum and natural-cone realizations with their corresponding modular conjugations. Section~\ref{sec:scalar} presents the explicit current model, including the global algebra, half-lines, bounded intervals, an explicit comparison with the canonical purification, and finally the other primary thermal states. To keep the main text focused, the review of the background material (operator algebras, states and their purifications, KMS states and modular theory) is deferred to Appendix~\ref{app:generalities}, and the properties of the natural positive cone are listed in Appendix~\ref{app:cone}.

\section{General setup,  strategy, and second worlds}\label{sec:setup}

This section fixes the notation and collects the few facts about purifications and modular theory that are used in the rest of the paper. All of them are standard \cite{haag1996local,BR1,BRII}. For the reader's convenience, a self-contained but non-exhaustive review is given in Appendix~\ref{app:generalities}, including the proofs of the simplest statements and the finite-dimensional examples that motivate them; the reader familiar with this material can safely skip it.

\paragraph{Algebras and states.} We work with von Neumann algebras $\mathcal{A}\subseteq\mathcal{B}(\mathcal{H})$, namely unital *-algebras of bounded operators equal to their bicommutant, $\mathcal{A}=\mathcal{A}''$, where the prime denotes the commutant. In QFT a von Neumann algebra $\mathcal{A}_{\mathcal O}$ is associated with every open spacetime region $\mathcal{O}$, and the C*-algebra generated by the algebras of all bounded regions is the quasi-local algebra (Appendix~\ref{app:algebras}). A state is a positive and normalized linear functional $\omega:\mathcal{A}\to\mathbb{C}$. It is faithful if $\omega(a^\dagger a)\neq 0$ for $a\neq 0$ and, when $\mathcal{A}$ is a von Neumann algebra, it is normal if it comes from a density matrix on $\mathcal{H}$, $\omega(a)=\Tr(\rho a)$ (Appendix~\ref{sec:states}). A state on the quasi-local algebra is locally normal if its restriction to every local algebra is normal. Thermal states are defined directly in infinite volume, where the Gibbs density matrix does not exist, as the states satisfying the KMS condition \eqref{kms} with respect to time translations (Appendix~\ref{app:thermal}).

\paragraph{Cyclic purifications.} A purification of a state $\omega$ on $\mathcal{A}$ is a triple $(\mathcal{K},\pi,|\psi\rangle)$ consisting of a Hilbert space $\mathcal{K}$, a representation $\pi:\mathcal{A}\to\mathcal{B}(\mathcal{K})$ and a vector $|\psi\rangle\in\mathcal{K}$ such that
\begin{equation}\label{setup:purification}
    \omega(a)=\langle\psi|\pi(a)|\psi\rangle,\qquad a\in\mathcal{A}.
\end{equation}
The purification is cyclic if $|\psi\rangle$ is cyclic for $\pi(\mathcal{A})$, i.e., if $\pi(\mathcal{A})|\psi\rangle$ is dense in $\mathcal{K}$. Every state admits a cyclic purification (the GNS construction), and any two cyclic purifications $(\mathcal{K}_1,\pi_1,|\psi_1\rangle)$ and $(\mathcal{K}_2,\pi_2,|\psi_2\rangle)$ of the same state are related by a unitary $U:\mathcal{K}_1\to\mathcal{K}_2$,
\begin{equation}\label{setup:unitary}
    \pi_2(a)=U\pi_1(a)U^\dagger,\qquad |\psi_2\rangle=U|\psi_1\rangle.
\end{equation}
If $\omega$ is a faithful normal state on a von Neumann algebra, then $\pi$ is faithful, $\pi(\mathcal{A})$ is again a von Neumann algebra, and $|\psi\rangle$ is also separating for it, i.e., $\pi(a)|\psi\rangle\neq0$ for $a\neq0$ (Appendix~\ref{sec:states}). In QFT, the Reeh-Schlieder theorem provides the basic example: the vacuum vector $|\Omega\rangle$ is cyclic and separating for the algebra of any non-empty region with non-empty causal complement.

\paragraph{Modular conjugation.} Let $(\mathcal{K},\pi,|\psi\rangle)$ be a cyclic purification of a faithful normal state $\omega$ on a von Neumann algebra $\mathcal{A}$. The Tomita operator is the closure of the antilinear operator defined by
\begin{equation}\label{setup:Tomita}
    S\pi(a)|\psi\rangle=\pi(a^\dagger)|\psi\rangle,\qquad S=J\Delta^{1/2},
\end{equation}
where the second equation is its polar decomposition. The modular operator $\Delta$ is positive and generates the modular group $\pi(a)\mapsto\Delta^{it}\pi(a)\Delta^{-it}$, which is the unique one-parameter group of automorphisms with respect to which $\omega$ satisfies the KMS condition at $\beta=-1$, that is, on the strip $-1\le\Im z\le0$ (Appendix~\ref{app:modular}). The \emph{modular conjugation} $J$ is antiunitary and satisfies $J^2=\mathds{1}$ and $J|\psi\rangle=|\psi\rangle$. The main result of Tomita-Takesaki theory states that
\begin{equation}\label{setup:TT}
    \Delta^{it}\pi(\mathcal{A})\Delta^{-it}=\pi(\mathcal{A}),\qquad J\pi(\mathcal{A})J=\pi(\mathcal{A})'.
\end{equation}
Both $\Delta$ and $J$ are uniquely determined by the algebra $\pi(\mathcal{A})$ and the cyclic and separating vector $|\psi\rangle$. Consequently, the modular conjugations of two cyclic purifications of the same state are related by the unitary in \eqref{setup:unitary},
\begin{equation}\label{setup:Jrelated}
    J_2=UJ_1U^\dagger.
\end{equation}
We will use this fact frequently: the state fixes $J$ only up to this unitary equivalence, and the concrete form of $J$, in particular where the operators $J\pi(a)J$ are localized, depends on the purification we choose. An equivalent definition of $J$ in terms of the analytic continuation of the modular group, which makes contact with expressions such as \eqref{intro:Tomita-vector}, as well as the explicit example of a density matrix purified in the space of Hilbert-Schmidt operators, can be found in Appendix~\ref{app:modular}.

\paragraph{Natural cone.} The closure $\mathcal{P}$ of the set of vectors $\pi(a)J\pi(a)J|\psi\rangle$, with $a\in\mathcal{A}$, is called the natural positive cone associated with $(\pi(\mathcal{A}),|\psi\rangle)$. We will need three of its properties \cite{Araki:1972zza,BR1}: every normal state on $\mathcal{A}$ is purified in the representation $\pi$ by a unique vector in $\mathcal{P}$; this vector is cyclic if the state is faithful; and the modular conjugation of any cyclic vector in $\mathcal{P}$ coincides with $J$. A more complete list of properties is given in Proposition~\ref{prop:cone} of Appendix~\ref{app:cone}.

\paragraph{Strategy.} In QFT, the Reeh-Schlieder theorem implies that the vacuum is cyclic and separating for the algebra ${\mathcal A}_{\mathcal O}$ associated with any non-empty region with non-empty causal complement. Thus, the vacuum defines both a faithful normal state on ${\mathcal A}_{\mathcal O}$ and a cyclic purification of that state. The corresponding modular group and modular conjugation are known for any QFT in the case where ${\mathcal O}$ is the Rindler wedge, $x>|t|$, where $x$ is a spatial coordinate and $t$ denotes time: the modular group assigns to each $s\in{\mathbb R}$ the boost of the $x$-$t$ plane with rapidity $-2\pi s$, and the modular conjugation is the CRT operator, where C stands for charge conjugation and RT is the transformation $(x,t)\mapsto -(x,t)$. This result is known as the Bisognano-Wichmann theorem \cite{Bisognano:1975ih,Bisognano:1976za} and will be exploited to obtain the modular conjugation for a KMS state  $\omega_\beta$ in different purifications $(\mathcal{K}, \pi,|\psi\rangle)$. However, we will have to take care of the fact that such result holds only for the vacuum state. The resolution is more or less well-known: consider a thermal state conformally related to the vacuum state (we will make this precise in the next Section). Then by conformal covariance of the theory one can map the vacuum $J_\Omega$ to the thermal one $J_\beta$. We will be able to also obtain $J$ for all the primary KMS states of the chiral boson, not only that conformally related to the vacuum. 

\paragraph{Second worlds.} Where does the second world enter? We give here a rough explanation and leave details for the next section. The easiest case to see where a second world appears is in the purification of the algebra of the whole space. Let us say that $\omega_\beta$ is a thermal state on the quasi-local algebra $\mathcal{A}_{\text{Mink}}$ of Minkowski spacetime and we consider a cyclic purification $(\mathcal{K}, \pi,|\psi\rangle)$ of such state. The corresponding modular conjugation $J_\beta$ maps fields in $\pi(\mathcal{A}_{\text{Mink}})\subset \mathcal{B}(\mathcal{K})$ to its commutant $\pi(\mathcal{A}_{\text{Mink}})'\subset \mathcal{B}(\mathcal{K})$. But there is no reason to expect that this commutant is in the image of $\pi$ , so the operators in the commutant do not generally correspond to quantum fields on Minkowski\footnote{Actually, $\pi(\mathcal{A}_{\text{Mink}})$ and $\pi(\mathcal{A}_{\text{Mink}})'$ are factors so they only intersect at multiples of the identity.}. A second world appears by forcing to view the operators in  $\pi(\mathcal{A}_{\text{Mink}})'$ as coming from a representation of quantum fields on Minkowski: there should be a second Minkowski spacetime algebra $\tilde{\mathcal{A}}_{\text{Mink}}={\mathcal{A}}_{\text{Mink}}$ to which we would apply $\pi$ to obtain the equality $\pi(\mathcal{A}_{\text{Mink}})'=\pi(\tilde{\mathcal{A}}_{\text{Mink}})$. {More precisely, since $a\mapsto J_\beta\pi(a)J_\beta$ is antilinear, what the commutant carries is a representation of the opposite algebra, $a^{\rm op}\mapsto J_\beta\pi(a^\dagger)J_\beta$. This is the right tensor factor in the doubled algebra $\mathcal{A}\otimes\mathcal{A}^{\rm op}$ of the canonical purification of \cite{sorce2025continuumcanonicalpurifications}, to which we come back at the end of Section~\ref{2dCFTs} and in Section~\ref{sec:sorce-comparison}.}

\section{2D CFTs and modular conjugation}\label{2dCFTs}

In this work we will mainly consider the thermal state given by the geometric KMS state which is obtained from the vacuum state by a conformal map. A rigorous and detailed construction of this geometric KMS state can be found in \cite{Camassa:2011te}, here we give a simpler description. Let us define the lightcone coordinates 
\begin{equation}\label{nullcoordinates}
    x^\pm=t\pm x_1 
\end{equation} 
and the conformal map
\begin{equation}\label{eq:sigmas}
\sigma(x^+,x^-)=(e^{\frac{2\pi}{\beta}x^+},-e^{-\frac{2\pi}{\beta}x^-})    
\end{equation}
Note that the image of {Minkowski} under $\sigma$ is precisely the right Rindler wedge $W_R=\left\{(t,x_1) \in \mathbb{R}^2\,;\, x_1>|t|\right\}$. A crucial and elementary property of $\sigma$ is that relates time translations with boosts: if $T_\tau(x^+,x^-)=(x^++\tau,x^-+\tau)$ is a time translation and $K_\tau(x^+,x^-)=(e^\tau x^+, e^{-\tau} x^-) $ a boost, then,
\begin{equation}
    \sigma \circ T_\tau = K_{\frac{2\pi}{\beta}\tau}\circ \sigma
\end{equation}
From now on, we work with one chirality and set $x=x^+$, so the whole spacetime means $x^+\in\mathbb{R}$ and the right Rindler wedge is given by $x^+>0$. The other chirality can be understood with little additional effort: for the right Rindler wedge we have $x^-<0$.

Now, we shall call ${\rm Exp}: \mathcal{A}_{\text{Mink}} \rightarrow \mathcal{A}_{\mathbb{R}_+}$ the conformal transformation\footnote{This  is defined in \cite{Camassa:2011te} and is shown it is an isomorphism of the quasi-local algebra of the real line $\mathcal{A}_{\mathbb{R}}$ to the quasi-local algebra of the positive real line $\mathcal{A}_{\mathbb{R}_+}$.} of the (primary) fields, induced by $\sigma$:
\begin{equation}
    {\rm Exp}(\phi(x)):=|\sigma'|^{h_+} \phi(\sigma(x))
\end{equation}
and define a new (geometric) state $\omega_\beta$ as
\begin{equation}
    \omega_\beta(\cdot):= \Omega({\rm Exp}(\cdot))
\end{equation}
Of course, both $\sigma$ and ${\rm Exp}$ depend on $\beta$ but we shall omit the subscript $\beta$. Notice that $\omega_\beta$ is the restriction of the vacuum state to the right Rindler wedge. We then have 
\begin{equation}
    \omega_\beta \left(\phi(T_\tau (x)) \right)=|\sigma'(x+\tau)|^{h_+}\Omega\left(\phi(\sigma(x+\tau))\right)=|\sigma'(x+\tau)|^{h_+}\Omega\left(\phi\left(K_{\frac{2\pi}{\beta}\tau}\circ\sigma(x)\right)\right)
\end{equation}
Namely, the state $\omega_\beta$ on a time translated operator by $\tau$ is equivalent to the vacuum state on a boosted operator by amount $\frac{2\pi}{\beta}\tau$. From Bisognano and Wichmann \cite{Bisognano:1976za}, we know that the modular group of the vacuum on $W_R$ is implemented by $\Delta^{it}_{\Omega,W_R}=U(K_{-2\pi t})$, the boost of parameter $-2\pi t$. Since the modular group satisfies the KMS condition at $\beta=-1$, the vacuum on $W_R$ is a KMS state at inverse temperature $2\pi$ with respect to the boosts $K_s$. So we get that $\omega_\beta$ satisfies the KMS property at inverse temperature $\beta$ with respect to translations on the whole Minkowski algebra, namely it is a thermal state and at temperature $\beta^{-1}$.    

\subsection{The conformal net on the plane}

We follow \cite{Tanimoto:2011pz,Morinelli:2018pof}. Let us consider the set of double cones $\mathcal{O}$ in two-dimensional Minkowski spacetime. A Haag-Kastler net is  a net of von Neumann algebras $\mathcal{A}_{\mathcal{O}}$ in $B(\mathcal{H})$  on a fixed Hilbert space $\mathcal{H}$, with a (strongly continuous) unitary  representation $U$ of the Poincaré group $\mathcal{P}_+^{\uparrow}$ and a vacuum vector $|\Omega\rangle$, such that the following axioms hold:
\begin{itemize}
    \item[] \textbf{Isotony:} if $\mathcal{O}_1 \subset \mathcal{O}_2$, then $\mathcal{A}_{\mathcal{O}_1}\subset \mathcal{A}_{\mathcal{O}_2}$
    \item[] \textbf{Locality:} if $\mathcal{O}_1$ is spacelike separated from  $\mathcal{O}_2$, then $\mathcal{A}_{\mathcal{O}_1}\subset \mathcal{A}_{\mathcal{O}_2}' $
    
    \item[] \textbf{Poincaré covariance:} $U(g)\mathcal{A}_{\mathcal{O}} U(g)^*=\mathcal{A}_{g\mathcal{O}}$ with $g\in \mathcal{P}_+^{\uparrow}$
    
    \item[] \textbf{Positivity of the energy:} the joint spectrum of the translation subgroup in $U$ is
contained in the closed forward light cone $\bar{V_+}=\left\{(p_0,p_1)\in\mathbb{R}^{1,1}\,,\,p_0\geq |p_1|\right\}$

\item[] \textbf{Existence of vacuum vector:} There is a unique (up to a phase) unit vector $|\Omega\rangle$ in $\mathcal{H}$ 
which is invariant under the action of $U$, and cyclic for the algebra generated by the union of all $\mathcal{A}_{\mathcal{O}}$. 

    \item[] \textbf{Additivity:} if $\mathcal{O}=\cup_i \mathcal{O}_i$, then $\mathcal{A}_\mathcal{O}=\bigvee_i\mathcal{A}_{\mathcal{O}_i}$
\end{itemize}
From these conditions, we get the Reeh-Schlieder property:
\begin{itemize}
    \item[] \textbf{Reeh-Schlieder:} The vector $|\Omega\rangle$ is cyclic and separating for each $\mathcal{A}_\mathcal{O}$.
\end{itemize}
The Bisognano-Wichmann {property} reads:
\begin{itemize}
    \item[] \textbf{Bisognano-Wichmann:} The modular group $\Delta^{it}$  of $\mathcal{A}_\mathcal{W_{R}}$ with respect to $|\Omega\rangle$ is equal to $U(\Lambda(-2\pi t))$, where $\Lambda(t)$ denotes the boost 
    $$\left(\begin{matrix}
\cosh t & \sinh t\\
\sinh t  & \cosh t        
    \end{matrix}\right)$$.
\end{itemize}
Although this property is not guaranteed by the previous axioms, if $U$ extends to the M\"obius group acting covariantly on the net extended to the Einstein cylinder, the net then is a M\"obius covariant net on the circle  and the Bisognano-Wichmann property is satisfied. Since we are interested in conformal field theories beyond the global conformal group, we have to ask for full diffeomorphism covariance on the lightlike coordinates $t\pm x$. In other words, we consider the chiral components of the M\"obius net, and define a conformal net on the circle and then restrict it to the real line where the lightlike coordinates belong. A (chiral) conformal net on the circle is  the same as the net described above, but: instead of double cones we have intervals $I \subset S^1$, the Poincaré group is replaced by the PSL$(2,\mathbb{R})$ group acting by {M\"obius} transformations on the circle, the positivity of energy is with respect to the generator of rotations of $S^1$, additivity is not an axiom but a consequence (see \cite{Tanimoto:2011pz} and references therein), and importantly we have local diffeomorphism covariance,

\begin{itemize}
    \item[] \textbf{Conformal covariance:} The representation $U$ extends to a projective unitary representation of Diff$(S^1)$ such that for any interval $I$ and $a \in \mathcal{A}_I$
    it holds that,
    \[U(g) \mathcal{A}_I U(g)^*=\mathcal{A}_{gI},\qquad g \in \text{Diff}(S^1)\]
    \[U(g) a U(g)^*=a,\qquad \text{supp}(g)\in I'\]
    
\end{itemize}
The circle can be viewed as the compactification of the real line by the Cayley transform, and the conformal net on the circle can be restricted to the real line $\mathcal{A}_\mathbb{R}$. Moreover, positivity of energy with respect to rotations of the circle is equivalent to positivity of the energy with respect to translations on the real line \cite{Longo:2008zz}. 

So we have the setup to study the modular conjugation operator of the thermal state $\omega_\beta$ on a chiral conformal net on the real line $\mathcal{A}_\mathbb{R}$. Note that, trivially, the vacuum state is purified by the vector $|\Omega\rangle$ with $\pi_\Omega=\text{id}$ on every $\mathcal{A}_{I}$. So from now on we will not make explicit $\pi_{\Omega}$ and just write $\pi_\Omega(a)$ as $a$ even if we talk about the vacuum representation.

\subsection{Purification of $\omega_\beta$ with vacuum vector}\label{globalpurifvacuumvector}

As we saw at the beginning of this section, we have a way of defining a thermal state $\omega_\beta$ as roughly the vacuum restricted to the right Rindler wedge. We can take this fact as a direct way to purify $\omega_\beta$ using the GNS representation of $\Omega|_{W_R}$ (the identity), namely we consider the triplet $(\mathcal{H},\pi_\beta,|\Omega\rangle)$ with:
\begin{equation}\label{pibetaglobal}
    \pi_\beta(\phi(x)):={\rm Exp}(\phi(x))
\end{equation}
Then, considering the $x^+$ chirality, for any field $\phi(x)$ localized anywhere in $\mathbb{R}$, 
\begin{equation}
   \langle \Omega|\pi_\beta(\phi(x))|\Omega\rangle=  \langle \Omega|{\rm Exp}(\phi(x))|\Omega\rangle=\Omega\left({\rm Exp}(\phi(x))\right)= \omega_\beta(\phi(x)) 
\end{equation}
By the Reeh-Schlieder theorem we know that $\Omega$ is cyclic and separating for $\mathcal{A}_{\mathbb{R}_+}$, so this is a cyclic representation for $\omega_\beta$.

From \cite{Bisognano:1976za} we also know\footnote{In the field formulas below we suppress any field-dependent
reflection or charge-conjugation action. Our
main purpose here is to track localization and give a consistent interpretation to fields evaluated at complex values. Except where displayed explicitly,
conformal Jacobian factors are also suppressed in the subsequent
point-field formulas. } the vacuum modular conjugation $J_{|\Omega\rangle,\mathcal{A}_{\mathbb{R}_+}}$ restricted to the algebra of observables $\mathcal{A}_{\mathbb{R}_+}$ of the wedge $W_R$: 
\begin{equation}
    J_{|\Omega\rangle,\mathcal{A}_{\mathbb{R}_+}} \phi(x) J_{|\Omega\rangle,\mathcal{A}_{\mathbb{R}_+}} = \phi(I_0(x)) =\phi(-x) ,\qquad x \in \mathbb{R}^+.
\end{equation}
Here $I_0(x)$ is the inversion with respect to the origin. Since the modular evolution and modular conjugation are uniquely defined by a von Neumann algebra and a cyclic and separating vector, we can claim that the modular conjugation  
\[J_{\beta,\mathbb{R}}:=J_{\beta,\pi_\beta(\mathcal{A}_{\mathbb{R}})}\] 
of the vector representative of $\omega_\beta$ (the vacuum vector $|\Omega\rangle$) over the whole algebra of observables is $J_{|\Omega\rangle,\mathcal{A}_{\mathbb{R}_+}}$:
\begin{equation}\label{Jglobal}
    J_{\beta,\mathbb{R}} \pi_\beta(\phi(x)) J_{\beta,\mathbb{R}} = |\sigma'(x)|^{h_+}\phi(-\sigma(x)) 
\end{equation}
Notice that $\phi(-\sigma(x))$ belongs to the algebra of the negative real line, or coming back to the two-dimensional Minkowski spacetime, to the algebra of the left Rindler wedge $W_L$ (actually, there are more possibilities in the 2D case which we discuss later).  We can rewrite this expression for the smeared field $\phi(f)$  by some function $f$ with arbitrary compact support (with the reflection action suppressed as above). In general,
\[
 {{\rm Exp}}
 (\phi(f))=\int_{\mathbb R_+}dy\,
 f(\sigma^{-1}(y))
 [\sigma'(\sigma^{-1}(y))]^{h_+-1}\phi(y).
\]
For weight $h_+=1$, the conformal factor cancels the Jacobian, and
we obtain:
\begin{equation}
    J_{\beta,\mathbb{R}} \pi_\beta(\phi(f)) J_{\beta,\mathbb{R}} = \phi(f\circ\sigma^{-1}\circ I_0)
\end{equation}
Notice that $f\circ\sigma^{-1}$ is supported on the positive real line, and therefore $\phi(f\circ\sigma^{-1}\circ I_0)$ belongs to the algebra of the negative real line. In other words, this field is not equal to some $\phi(g \circ \sigma^{-1})$ for some appropriate function $g$ and therefore it does not belong to the representation $\pi_\beta$. The commutant of the algebra generated by $\pi_\beta(\phi(f))$ lives outside the representation, and it should not come as a surprise: there are no fields outside Minkowski. However the commutant is not empty since \textit{there are operators} commuting with the algebra generated by  $\pi_\beta(\phi(f))$, these are the vacuum fields smeared with functions with support on the negative real line. 

\subsection{Purification of ${\omega_\beta}|_{\mathcal{A}_\mathcal{O}}$ with vacuum vector}\label{purifvacuumvector}

What about the thermal state restricted to some wedge $W_{r}$, the wedge $W_R$ spacelike translated by $r$? So we make room for other fields of the type $\pi_\beta(\phi(f))$ located outside $W_{r}$. In particular we expect the commutant of $\mathcal{A}_{W_{r}}$, $\mathcal{A}_{W_{r}}'$, contains elements like $\pi_\beta(\phi(f))$ with $f$ supported inside $W_{r}'$, the causal complement of $W_{r}$. We are calling $W_{r}$ the right wedge with edge at $x=r$, which in the positive chirality corresponds to $x^+>r$ (and $x^-<-r$ for the other chirality). In order to avoid cluttering of the notation, we shall write $\mathcal{A}_r$ for the algebra associated with the interval $(r,\infty)$. With this notation, $\mathcal{A}_{0}$ is the algebra of the positive real line. We could consider the local algebra associated to a large double cone $\mathcal{A}_\mathcal{O}$ instead of $\mathcal{A}_r$, and the idea would be exactly the same, so we just stick to wedges for now (see next Subsection).

In order to study this, we can take the previous purification by means of the cyclic representation \eqref{pibetaglobal}, and then restrict this representation to $\mathcal{A}_{r}$, or we can first restrict the algebraic state $\omega_\beta$, obtain $\omega_\beta|_{\mathcal{A}_{r}}$, and then choose some purification (which may or {may} not coincide with the previous one). In the latter case we can in fact use the vacuum representation, which we will do later. For now, let us take \eqref{pibetaglobal} and restrict it to fields $\phi(x)$ such that $x>r$, for some finite $r\in\mathbb{R}$. The difference with the previous case where we considered the algebra of the real line, is that now $\sigma$ maps $(r,\infty)$ to $(\sigma
(r),\infty)$, so we do not have the result of Bisognano and Wichmann at hand valid for $\mathbb{R}_+$. Of course, it is simple to adapt, by translating $(\sigma
(r),\infty)$ to the positive real line, applying $I_0$, and translating back to  $(\sigma
(r),\infty)$. The claim is that 
\begin{equation}
J_{|\Omega\rangle,\mathcal{A}_{\sigma(r)}}=U_{T_{\sigma(r)}} J_{|\Omega\rangle,\mathcal{A}_{\mathbb{R}_+}} U_{T_{\sigma(r)}}^\dagger
\end{equation}
which follows from the definition of the Tomita operator $S$ (for each local algebra) and covariance of the net of algebras. We then have 
\begin{equation}
    J_{\beta,\mathcal{A}_{r}} =J_{|\Omega\rangle,\mathcal{A}_{\sigma(r)}}=U_{T_{\sigma(r)}} J_{|\Omega\rangle,\mathcal{A}_{\mathbb{R}_+}} U_{T_{\sigma(r)}}^\dagger
\end{equation}
Explicitly, 
\begin{equation}
    J_{|\Omega\rangle,\mathcal{A}_{\sigma(r)}} \phi(\sigma(x)) J_{|\Omega\rangle,\mathcal{A}_{\sigma(r)}} = \phi(2\sigma(r)-\sigma(x)) ,\qquad x \in (r,\infty).
\end{equation}
Unsurprisingly, $\sigma(x)\mapsto 2\sigma(r)-\sigma(x)$ is a reflection with respect to $\sigma(r)$, which we shall call $I_{\sigma(r)}$. This means that $\phi(2\sigma(r)-\sigma(x))$ belongs to $\mathcal{A}_{\sigma(r)}'=\mathcal A_{(-\infty,\sigma(r))}$. Notice that if $2\sigma(r)-\sigma(x)>0$, the field  $\phi(2\sigma(r)-\sigma(x))$ is of the form $\pi_\beta(\phi(y))$, with $y=\sigma^{-1}(2\sigma(r)-\sigma(x))$, so it belongs to the representation $\pi_\beta$.  This is different from the previous situation of  the modular conjugation of the global algebra, where there was no room in the representation $\pi_\beta$ for the commutant of $\mathcal{A}_{\mathbb{R}_+}$. By restricting to some smaller local algebra such as $\mathcal{A}_r$, the commutant $\pi_\beta(\mathcal{A}_r)'=\mathcal{A}_{\sigma(r)}'=\mathcal{A}_{(-\infty,\sigma(r))}  $ is larger than before and some elements of it belong to $\pi_\beta(\mathcal{A}_{\mathbb{R}})=\mathcal{A}_{0}$, which in 2D would be the right Rindler Wedge. 

\textcolor{black}{ Let us discuss in more detail how to translate the previous 1D analysis to the 2D spacetime. Even though  the 2D theory is chiral, and the 2D algebra is ``simply'' the tensor product of the chiral algebras, there is an important difference: in 1D the commutant of the algebra of an interval $\mathcal{I}$ is the algebra of the complement $\mathcal{I}'$\footnote{For a half-line this is Haag duality on the circle, $\mathcal{A}_{\mathcal{I}}'=\mathcal{A}_{\mathcal{I}'}$, which holds for every conformal net: the complement of a half-line is the other half-line. For a bounded interval, $\mathcal{I}'$ is the complementary interval of the circle, which contains the point at infinity, and its algebra coincides with the algebra generated by the two complementary half-lines of the real line if and only if the net is strongly additive \cite[Lemma~1.3]{GuidoLongoWiesbrock1998}. This is the case for the $U(1)$-current net \cite[Corollary~2.11]{GuidoLongoWiesbrock1998}, but Haag duality on the real line may fail for a general conformal net.}, but in 2D the commutant of the algebra of a wedge region is not the algebra of the complement region, but the algebra of the causal complement, which is smaller. In other words, regions such as the future or past light cones have to be considered.}

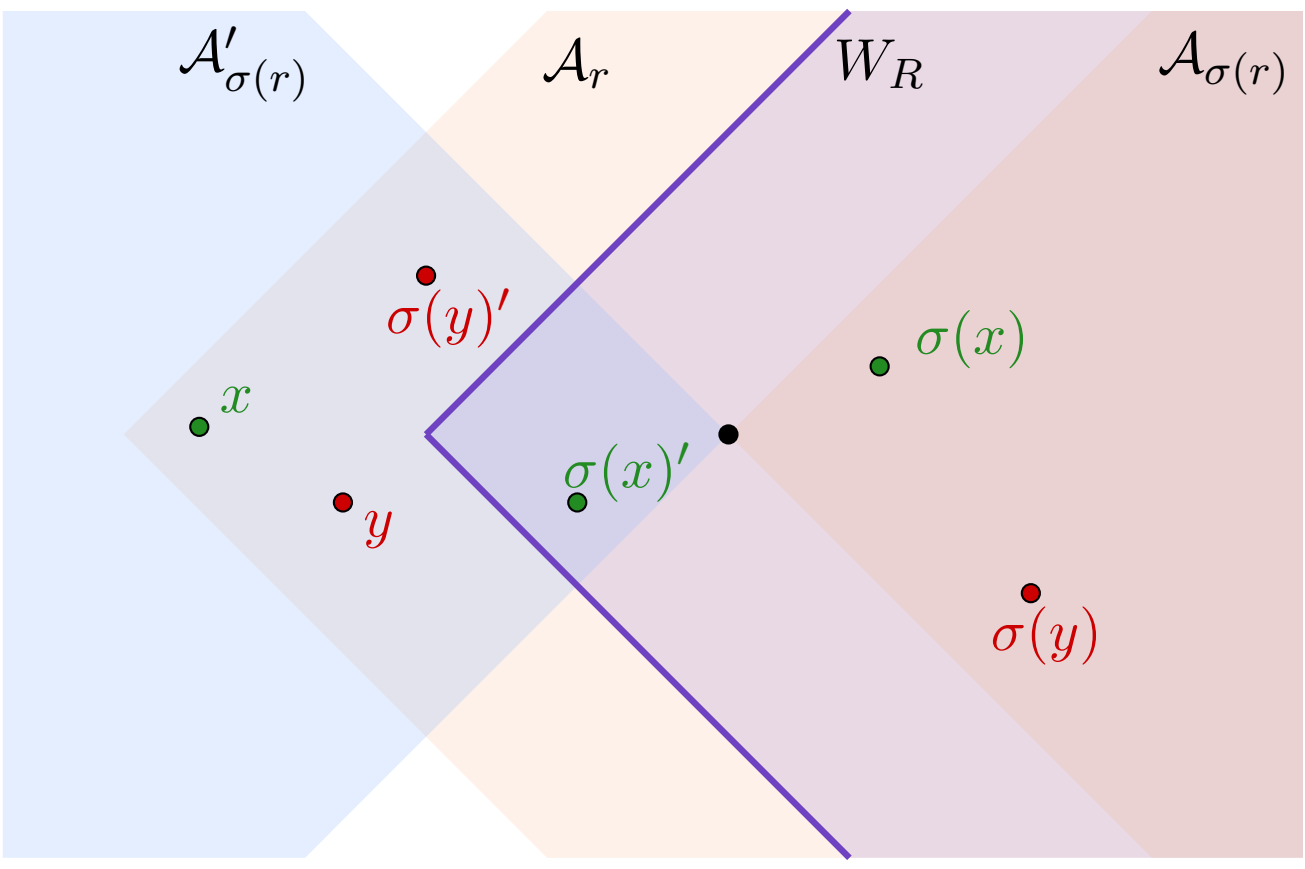
\begin{figure}[h!]
\centering
\begin{tikzpicture}[scale=1]


\def\a{2}
\def\b{4}
\def\Xmax{6.8}
\def\Xmin{2.8}
\def\Tmax{1}


\begin{scope}
\clip
  (0,0)
  --
  (\Xmin,\Xmin)
  --
  (\Xmax+\Tmax,\Xmin)
  --
  (\Xmax+\Tmax,-\Xmin)
  --
  (\Xmin,-\Xmin)
  --
  cycle;

\fill[AsFill,opacity=.18]
  (-1,-\Xmax)
  rectangle
  (\Xmax+\Tmax,\Xmax);
\end{scope}


\begin{scope}
\clip
  (\a,0)
  --
 (\Xmin+\a,\Xmin)
  --
  (\Xmax+\Tmax,\Xmin)
  --
  (\Xmax+\Tmax,-\Xmin)
  --
  (\Xmin+\a,-\Xmin)
  --
  cycle;

\fill[ArFill,opacity=.18]
  (\a-1,-\Xmax)
  rectangle
  (\Xmax+\Tmax,\Xmax);
\end{scope}


\begin{scope}
\clip
  (\b,0)
  --
  (\Xmin+\b,\Xmin)
  --
  (\Xmax+\Tmax,\Xmin)
  --
  (\Xmax+\Tmax,-\Xmin)
  --
  (\Xmin+\b,-\Xmin)
  --
  cycle;

\fill[AsFill,opacity=.18]
  (\b-1,-\Xmax)
  rectangle
  (\Xmax+\Tmax,\Xmax);
\end{scope}


\begin{scope}
\clip
  (\b,0)
  --
  (\b-\Xmin, \Xmin)
  --
  (-\Xmin+\a, \Xmin)
  --  
  (-\Xmin+\a,-\Xmin)
  --
  (\b-\Xmin, -\Xmin)
  --
  cycle;

\fill[AspFill,opacity=.23]
  (\Xmin+\b,\Xmin)
  rectangle
  (-\Xmin,-\Xmin);
\end{scope}






\draw[
  ArLine,
  line width=1.2pt
]
  (\a,0)
  --
  (\Xmin+\a,\Xmin);

\draw[
  ArLine,
  line width=1.2pt
]
  (\a,0)
  --
  (\Xmin+\a,-\Xmin);







\coordinate (x)   at (0.5,0.05);
\coordinate (sx)  at (5,0.45);
\coordinate (sxp) at (3,-0.45);


\filldraw[
  fill=ForestGreen,
  draw=black
]
  (x)
  circle (1.7pt);

\filldraw[
  fill=ForestGreen,
  draw=black
]
  (sx)
  circle (1.7pt);

\filldraw[
  fill=ForestGreen,
  draw=black
]
  (sxp)
  circle (1.7pt);

\node[
  ForestGreen,
  anchor=west
]
  at ($(x)+(0,0.18)$)
  {$x$};

\node[
  ForestGreen,
  anchor=west
]
  at ($(sx)+(0.1,0.18)$)
  {$\sigma(x)$};

\node[
  ForestGreen,
  anchor=west
]
  at ($(sxp)-(0.23,-0.2)$)
  {$\sigma(x)'$};


\coordinate (y)   at (1.4
5,-0.45);
\coordinate (sy)  at (6,-1.05);
\coordinate (syp) at (2,1.05);
\coordinate (Is) at (4,0);


\filldraw[
  fill=black!80!black,
  draw=black
]
  (Is)
  circle (1.7pt);

\filldraw[
  fill=red!80!black,
  draw=black
]
  (y)
  circle (1.7pt);

\filldraw[
  fill=red!80!black,
  draw=black
]
  (sy)
  circle (1.7pt);

\filldraw[
  fill=red!80!black,
  draw=black
]
  (syp)
  circle (1.7pt);

\node[
  red!80!black,
  anchor=west
]
  at ($(y)+(0,-0.18)$)
  {$y$};

\node[
  red!80!black,
  anchor=west
]
  at ($(sy)+(-.4,-0.28)$)
  {$\sigma(y)$};

\node[
  red!80!black,
  anchor=west
]
  at ($(syp)+(-.4,-0.28)$)
  {$\sigma(y)'$};


\node at (3,2.45)
  {$\mathcal A_{r}$};

\node at (5,2.45)
  {${W}_R$};

\node at (7.28,2.45)
  {$\mathcal A_{\sigma(r)}$};

\node at (.8,2.45)
  {$\mathcal A'_{\sigma(r)}$};

\end{tikzpicture}
\caption{The (algebra associated to the) wedge $\mathcal{A}_r$, its conformal image
$\mathcal A_{\sigma(r)}$, the reflected wedge
$\mathcal A'_{\sigma(r)}$ with respect to {the} edge $(0,\sigma(r))$ (black dot), and the right Rindler wedge ${W}_R$. The latter has null boundaries in solid purple lines, and to their right we have the purification $\pi_\beta$. We see that a field located at $x$ is represented at $\sigma(x)$ and the modular conjugation sends to $\phi(\sigma(x)')$, which belongs to the purification. On the other hand,  a field at $y$ is represented at $\sigma(y)$ and the modular conjugation sends to $\phi(\sigma(y)')\notin \mathcal{A}_{{W}_R}$, which is not in the purification (here it is drawn in the future light cone).} 
\label{fig:vacpurification}
\end{figure}

In more precise terms, distinguishing both chiralities, we have,
\begin{equation}
\mathcal{A}_{{W}_r}=\mathcal{A}_r^+ \otimes (\mathcal{A}_{-r}^-)'
\end{equation}
where $(\mathcal{A}_{-r}^-)'$ is the algebra of the region $(-\infty,-r)$ for the negative chirality. The thermal purification we discussed before now reads,
\begin{equation}
    \pi_\beta(\mathcal{A}_{{W}_r})=\mathcal{A}_{{W}_{\sigma(r)}}=\mathcal{A}_{\sigma(r)}^+ \otimes (\mathcal{A}_{-\sigma(r)}^-)',\qquad \sigma(r)=e^{\frac{2\pi}{\beta}r}
\end{equation}
Calling $v=x^+$ and $u=x^-$, we have,
\begin{equation}
    J_{\beta,\mathcal{A}_r}\pi_\beta(\phi(u,v))J_{\beta,\mathcal{A}_r}=\phi(-2\sigma(r)-\sigma_-(u),2\sigma(r)-\sigma_+(v))
\end{equation}
where $\sigma_\pm$ are read from \eqref{eq:sigmas}. As opposed to the 1D case where there were two possibilities ($2\sigma(r)-\sigma_+(v)\lessgtr 0$ ), now we have four. Let us call $\tilde{\sigma}_+(v)=2\sigma(r)-\sigma_+(v)$ and $\tilde \sigma_-(u)=-2\sigma(r)-\sigma_-(u)$ in order to simplify the notation. Then, 
\begin{enumerate}
    \item[i] If $\tilde\sigma_+(v)>0$ and $\tilde\sigma_-(u)<0$, the modular conjugation maps $\pi_\beta(\phi(u,v))$ to another operator in the representation, namely to $\pi_\beta(\phi(\sigma_-^{-1}\circ\tilde\sigma_-(u),\sigma_+^{-1}\circ\tilde\sigma_+(v)))$, which belongs to the algebra $\mathcal{A}_{W_R}$.\\
    \item[ii] If $\tilde\sigma_+(v)<0$ and $\tilde\sigma_-(u)>
0$, the modular conjugation maps to the algebra  $\mathcal{A}_{W_L}$, of the left Rindler Wedge.\\
\item[iii] If $\tilde\sigma_+(v) \tilde\sigma_-(u)>0$, the modular conjugation maps to the algebra of the future (both positive) or past (both negative) light cones.  
\end{enumerate}
Only for the first case the modular conjugation maps to an operator in the purification. Otherwise, we have a field $\phi(\tilde\sigma_-(u),\tilde\sigma_+(v))$ outside the representation, but is hard to give it an interpretation of  ``second world'' since there are three other regions, not only the causal complement $W_L$ of $W_R$ (see Figure \ref{fig:vacpurification}). However, {maybe} the best ``second world'' naming should be assigned to the commutant of $\mathcal{A}_{W_{\sigma(r)}}$, where all these cases belong. The problem with this is that the second world moves as we move $r$. Also note that case ii above, where the modular conjugated field belongs to $\mathcal{A}_{W_L}$, corresponds to the second world interpretation of \cite{Mintchev:2022fcp}, where that wedge is described with null coordinates in the whole real line and therefore interpreted as a second Minkowski spacetime.

Coming back to the positive chiral part, for a weight-one smeared field, and still suppressing its reflection parity, we obtain  
\begin{equation}\label{Jhalfline}
    J_{\beta,\mathcal{A}_{r}} \pi_\beta(\phi(f)) J_{\beta,\mathcal{A}_{r}} = \phi(f\circ\sigma^{-1}\circ T_{\sigma(r)} \circ I_0 \circ T_{-\sigma(r)}),\qquad \text{supp}(f)\subset (r,\infty)
\end{equation}
As expected,  we obtain a transformed field smeared over a function with support in $(-\infty,\sigma(r))$. The ``second world'' may appear if one insists that all the fields like \eqref{Jhalfline} are still in the image of $\pi_\beta$, that is the RHS of \eqref{Jhalfline} be of the form $\phi(g\circ\sigma^{-1})$, for some compactly supported function $g$. However, by the previous discussion, we know that only when the field $\phi(f)$ is localized inside the interval $(r,\sigma^{-1}(2\sigma(r)))$ this can happen, otherwise the (representation of the) field transformed by the modular conjugation ``leaks''  to the negative real line and there is no way it is of the form $\phi(g\circ\sigma^{-1})=\pi_\beta(\phi(g))$, which can only be localized in the positive real line. It is when we insist that the transformed field matches  $\phi(g\circ\sigma^{-1})$ that there is some trouble and $g\circ\sigma^{-1}$ will be evaluated at a negative number. This is the root of the complex numbers  appearing sometimes in the literature as $\phi(x\pm i\frac{\beta}{2} )$ \cite{Mintchev:2022fcp,AbateKoifman2023}. 

Another way the ``second world'' may appear is in a purification where we duplicate the Hilbert space $\mathcal{H}$, as in type I algebras, and this is what we do in the next Section for the special case of a free field. But first, let us discuss another way to purify $\omega_\beta$ restricted to some local algebra. We have already done this using \eqref{pibetaglobal} restricted to a local algebra, but now we want to use another purification, one where the representation is the identity, namely that of the vacuum. The trade-off is that the vector representative is not the vacuum vector $|\Omega\rangle$ anymore.

\subsection{Purification of $\omega_\beta\Big|_{\mathcal{A}_{\mathcal{O}}}$ with natural cone vector}\label{sec:positivecone}

First some technical remarks: we will need that the KMS states are locally normal in order to use a result about the modular conjugation in the context of natural cones. Local normality means that the state restricted to the local algebra $\mathcal{A}_\mathcal{O}$ is normal, namely it can be written as a (non-unique) density matrix acting on $\mathcal{H}$. We are not aware of any result that guarantees local normality of a KMS state with some generality for a net of local von Neumann algebras. Even for the more stringent case of a {blue}{M\"obius} net on the circle restricted to the real line $\mathcal{A}_\mathbb{R}$ there are no known nice sufficient conditions. Actually, there are cases where a KMS is not locally normal, for example a free massless scalar field in three spacetime dimensions \cite{Buchholz}. However, if the net is generated by local von Neumann algebras with the split property, by \cite{Takesaki:1973ge}, local normality of KMS states follows\footnote{We are grateful to Yoh Tanimoto for correspondence about these technical points.}. Moreover, the split property is automatic for conformal nets on the circle \cite{Morinelli:2016wcz}. We can then obtain a conformal net on 2D with the split property by taking the product of two chiral conformal nets with the split property. We are assuming then that the models under consideration satisfy the split property (for example the $U(1)$ current net describing a chiral boson or the net Vir${}_c$ generated by stress-energy tensors, see \cite{Camassa:2011te,Camassa:2011wk} for further details).

We want to purify the thermal state restricted to some vacuum von Neumann local algebra, $\mathcal{A}_{\mathcal{O}}$, without invoking another local algebra. In other words, we want that the thermal cyclic representation of $\mathcal{A}_{\mathcal{O}}$ be the identity. To be more precise, consider some big enough double cone  $\mathcal{O}_{\text{Big}}$ that contains $\mathcal{O}$ and let us call its vacuum algebra $\mathcal{A}_{\text{Big}}$, so the vacuum state $|\Omega\rangle$ is a cyclic and separating vector for $\mathcal{A}_{\text{Big}}$. Let us then consider the corresponding natural cone $\mathcal{P}_{\text{Big}}$, 
\begin{equation}
    \mathcal{P}_{\text{Big}}=\overline{\left\{\Delta^{1/4}a|\Omega\rangle ;\qquad a \in \mathcal{A}_{\text{Big}}, \, a\geq 0 \right\}}
\end{equation}
Then we consider the restriction of $\omega_\beta$ to $\mathcal{A}_{\text{Big}}$. By property 8 of Proposition~\ref{prop:cone}, for a (locally normal) KMS state $\omega_\beta$ restricted to $\mathcal{A}_{\text{Big}}$, $\omega_\beta|_{\mathcal{A}_{\text{Big}}}$, we have a vector state $|\Psi_{\text{Big}}\rangle\in \mathcal{P}_{\text{Big}}$ such that
\begin{equation}
    \langle \Psi_{\text{Big}}|a|\Psi_{\text{Big}}\rangle=\omega_\beta|_{\mathcal{A}_{\text{Big}}}(a),\qquad a \in \mathcal{A}_{\text{Big}}
\end{equation}
So we have achieved a cyclic purification of ${\omega_\beta}|_{\mathcal{A}_{\text{Big}}}$ in the vacuum representation Hilbert space $\mathcal{H}$ with representation given by the identity (the representation is cyclic because the state is faithful on $\mathcal{A}_{\text{Big}}$ \footnote{Any KMS state is faithful on $\mathcal{A}_{\text{Big}}$ if the quasi-local algebra is generated by factors: this is because on each factor the GNS representation of the state $\omega_\beta$ is faithful (namely, injective). This and the separating property of the GNS vector, which follows by an argument as in Corollary 5.3.9 of \cite{BRII}, gives a faithful state on $\mathcal{A}_{\text{Big}}$.} which implies any vector representative is  separating, which by property 6 of Proposition~\ref{prop:cone} implies cyclic). Even more, the state representative lives in the self-dual cone $\mathcal{P}_{\text{Big}}$ associated with the algebra $\mathcal{A}_{\text{Big}}$. In short, the representation is 
\begin{equation}\label{tripletcone}
    (\mathcal{H},\, \pi_\Omega=\text{id},\,|\Psi_{\text{Big}}\rangle\in\mathcal{P}_{\text{Big}} )
\end{equation} 
With respect to the purification of the previous subsection, we have somewhat simplified the representation changing $\pi_\beta= {\rm Exp}$ for $\pi_\Omega$ which is the identity, with the price to pay that now the vector representative is no longer the vacuum vector $|\Omega\rangle$ but some vector $\Psi_{\text{Big}}\in\mathcal{P}_{\text{Big}}$.

Now, in order to compute the modular conjugation associated to the subalgebra algebra ${\mathcal{A}_{\mathcal{O}}\subset \mathcal{A}_{\text{Big}}}$, we proceed similarly as in the previous section. First of all, let us consider a chiral component  so we have a conformal net on the circle restricted to the real line. So we change notation accordingly with respect to the previous lines: now we have a big enough interval $\mathcal{I}_{\text{Big}}$ and repeating all the reasoning above, the cyclic representation of the local thermal state is \eqref{tripletcone} with the cone $\mathcal{P}_{\text{Big}}$ associated to the algebra of a big interval 
\begin{equation}
\mathcal{A}_{\text{Big}}:=\mathcal{A}_{\mathcal{I}_{\text{Big}}}.
\end{equation} 
We are looking for the modular conjugation $J_{\mathcal{I},\beta}$ associated to $\omega_\beta|_{\mathcal{I}}$ and the algebra $\mathcal{A}_{\mathcal{I}}\subseteq \mathcal{A}_{\text{Big}}$, with $\mathcal{I}\subseteq{\mathcal{I}}_{\text{Big}}$.

Let us define $U_{{\text{Big}}}:\mathcal{H}\rightarrow \mathcal{H}$,
\begin{equation}
    U_{{\text{Big}}}\pi_\beta(a)|\Omega\rangle=U_{{\text{Big}}} {\rm Exp}(a)|\Omega\rangle=a|\Psi_{{\text{Big}}}\rangle,\qquad a \in A_{{\text{Big}}}
\end{equation}
Note that this could not be possible if $\mathcal{A}_{{\text{Big}}}$ is taken to be the whole real line, since in that case $|\Psi_{{\text{Big}}\rangle}$ is not separating (the complement of the real line is empty). This operator $U_{{\text{Big}}}$ is densely defined and an isometry, so it can be extended to a unitary operator on $\mathcal{H}$ which we continue to call $U_{{\text{Big}}}$. In particular, we can map on a dense domain one cyclic representation to the other,
\begin{equation}\label{representationsiso}
    U_{{\text{Big}}}\pi_\beta(a) U_{{\text{Big}}}^\dagger b |\Psi_{{\text{Big}}}\rangle=U_{{\text{Big}}}\pi_\beta(a) \pi_\beta(b) |\Omega\rangle=ab|\Psi_{{\text{Big}}}\rangle
\end{equation}
showing that indeed, 
\begin{equation}\label{eq:Ubig}
    U_{{\text{Big}}}\pi_\beta(a) U_{{\text{Big}}}^\dagger=a, \qquad a\in\mathcal{A}_{{\text{Big}}}.
\end{equation}
Note that for a local algebra $\mathcal{A}_I\subseteq \mathcal{A}_{{\text{Big}}}$ the reduced state $\omega_\beta|_{\mathcal{A}_I}$ is still represented by $|\Psi_{{\text{Big}}}\rangle$:
\begin{equation}\label{conerepresentative}
    \omega_{\beta}|_{\mathcal{A}_I}(a)=(|\Psi_{{\text{Big}}}\rangle,a|\Psi_{{\text{Big}}}\rangle), \qquad a\in\mathcal{A}_I
\end{equation}
but this is as much as we can say, as we do not know how to relate cones and inclusions of subalgebras. In any case, having the cyclic representation \eqref{tripletcone}, and \eqref{conerepresentative}, together with the unitary $U_{{\text{Big}}}$ will be enough for our purposes. The subtle point is that even though $|\Psi_{\text{Big}}\rangle$ is cyclic for $\mathcal{A}_{\text{Big}}$, it is not so clear that it is also cyclic for $\mathcal{A}_{\mathcal{I}}$, and we need this in order to have $J_{|\Psi_{\text{Big}}\rangle,\mathcal{I}}$ well defined\footnote{Actually, there is a modular theory for non-cyclic non-separating vectors, as nicely reviewed in the Appendix of \cite{ceyhan2020recovering}.  }. Note that \eqref{tripletcone} is a cyclic representation for the big algebra $\mathcal{A}_{\text{Big}}$,  but it may not work as a \text{cyclic} representation for  $\mathcal{A}_{\mathcal{I}}$, since the cone changes as we change the algebra, and so does the vector representative $|\Psi\rangle$ in the corresponding cone. Despite these words of caution, it turns out that $|\Psi_{\text{Big}}\rangle$ is indeed a cyclic vector for $\mathcal{A}_{\mathcal{I}}$: for any $a\in \mathcal{A}_{\mathcal{I}}$,  \begin{align}
    a|\Psi_{\text{Big}}\rangle&=aU_{\text{Big}}|\Omega\rangle\nonumber\\
    &=U_{\text{Big}}U_{\text{Big}}^\dagger aU_{\text{Big}}|\Omega\rangle\nonumber\\
    &=U_{\text{Big}}\pi_\beta(a)|\Omega\rangle\nonumber\\
    &=U_{\text{Big}}{\rm Exp}(a)|\Omega\rangle\nonumber.
\end{align}
Since ${\rm Exp}(a)$ belongs to $\mathcal{A}_{\sigma(\mathcal{I})}$ which is the local algebra of another interval, by Reeh-Schlieder ${\rm Exp}(\mathcal{A}_{\mathcal{I}})|\Omega\rangle$ is dense in $\mathcal{H}$. This means that $U_{\text{Big}}{\rm Exp}(\mathcal{A}_{\mathcal{I}})|\Omega\rangle$ is also dense, and so is $\mathcal{A}_{\mathcal{I}}|\Psi_{\text{Big}}\rangle$.

In order to obtain the modular conjugation for $\mathcal{A}_{\mathcal{I}}$ and $|\Psi_{\text{Big}}\rangle$, we again use covariance of the net and the existence of $U_{\text{Big}}$ relating two cyclic representations,
\begin{equation}\label{eq:Jsrelated}
    J_{|\Psi_{\text{Big}}\rangle,\mathcal{I}}=U_{\text{Big}} J_{|\Omega\rangle,\sigma(\mathcal{I})}U_{\text{Big}}^\dagger.
\end{equation}
Now, the inversion of an interval $\mathcal{I}=(a,b)$  is the geometric action of the modular conjugation $J_{|\Omega\rangle,\mathcal{I}}$ of the vacuum, and is given by,
\begin{equation}
    I_{(a,b)}(x)=\frac{cx-ab}{x-c},\qquad c=\frac{a+b}{2}
\end{equation}
For the following point-field formulas we suppress the conformal
Jacobian factors as well as the reflection action, since only the
localization of the transformed operator is relevant. Taking this into
account, we arrive at
\begin{equation}
   J_{|\Psi_{\text{Big}}\rangle,\mathcal{I}}\,\, \phi(x)\,\, J_{|\Psi_{\text{Big}}\rangle,\mathcal{I}} = U_{\text{Big}} \phi\left(I_{\sigma(\mathcal{I})}\circ\sigma(x)\right)U_{\text{Big}} ^\dagger
\end{equation}
If $I_{\sigma(\mathcal{I})}\circ\sigma(x) \in \sigma(\mathcal{I}_\text{Big})$, by \eqref{eq:Ubig} we can say the RHS gives $\phi\left(\sigma^{-1}\circ I_{\sigma(\mathcal{I})}\circ \sigma (x)\right)\in\mathcal{A}_\text{Big}$. However, if $I_{\sigma(\mathcal{I})}\circ\sigma(x) \notin \sigma(\mathcal{I}_\text{Big})$ we cannot use \eqref{eq:Ubig}, but we can proceed as follows,
\begin{align}
    J_{|\Psi_{\text{Big}}\rangle,\mathcal{I}}\,\, \phi(x)\,\, J_{|\Psi_{\text{Big}}\rangle,\mathcal{I}} 
    &=U_{\text{Big}} J_{|\Omega\rangle,\sigma(\mathcal{I}_\text{Big})}\,  \phi\left(I_{\sigma(\mathcal{I}_\text{Big})}\circ I_{\sigma(\mathcal{I})}\circ\sigma(x)\right)J_{|\Omega\rangle,\sigma(\mathcal{I}_\text{Big})}\,U_{\text{Big}} ^\dagger\nonumber\\
    &=U_{\text{Big}} J_{|\Omega\rangle,\sigma(\mathcal{I}_\text{Big})}\,  \,U_{\text{Big}} ^\dagger\phi\left(\sigma^{-1}\circ I_{\sigma(\mathcal{I}_\text{Big})}\circ I_{\sigma(\mathcal{I})}\circ\sigma(x)\right)\,U_{\text{Big}}\, J_{|\Omega\rangle,\sigma(\mathcal{I}_\text{Big})}\,U_{\text{Big}} ^\dagger\nonumber\\
    &= J_{|\Psi_{\text{Big}}\rangle,\mathcal{I}_\text{Big}}\,\phi\left(\sigma^{-1}\circ I_{\sigma(\mathcal{I}_\text{Big})}\circ I_{\sigma(\mathcal{I})}\circ\sigma(x)\right)\,J_{|\Psi_{\text{Big}}\rangle,\mathcal{I}_\text{Big}}\nonumber
\end{align}
where we used \eqref{eq:Ubig} in the second line and  \eqref{eq:Jsrelated} in the third line. 
Finally, by property 7 of Proposition~\ref{prop:cone}, we know that $J_{|\Psi_{\text{Big}}\rangle,\mathcal{I}_{\text{Big}}}=J_{|\Omega\rangle,\mathcal{I}_{\text{Big}}}$, so we can write
\begin{align}\label{eq:Jactioncone}
    J_{|\Psi_{\text{Big}}\rangle,\mathcal{I}}\,\, &\phi(x)\,\, J_{|\Psi_{\text{Big}}\rangle,\mathcal{I}} =\nonumber\\
    &=\left\{ 
\begin{array}{ccc}
         \phi\left( \sigma^{-1}\circ I_{\sigma(\mathcal{I})}\circ \sigma (x)\right), & &I_{\sigma(\mathcal{I})}\circ\sigma(x)\in \sigma(\mathcal{I}_{\text{Big}})\\
    \phi\left(I_{\mathcal{I}_{\text{Big}}}\circ\sigma^{-1}\circ I_{\sigma(\mathcal{I}_{\text{Big}})}\circ I_{\sigma(\mathcal{I})}\circ \sigma (x)\right),&& I_{\sigma(\mathcal{I})}\circ\sigma(x)\in \sigma(\mathcal{I}_{\text{Big}})'
    \end{array}    
    \right.
\end{align}
Note that for the first case in \eqref{eq:Jactioncone}, we see that $I_{\sigma(\mathcal{I})}\circ \sigma (x)\in\left\{\mathbb{R}-\sigma(\mathcal{I})\right\} \cap \sigma(\mathcal I_\text{Big}) $, so $\sigma^{-1}\circ I_{\sigma(\mathcal{I})}\circ \sigma (x) \in \mathbb{R}-\mathcal I$ and $\phi(\sigma^{-1}\circ I_{\sigma(\mathcal{I})}\circ \sigma (x))\in \mathcal A _{\mathcal{I}}'$. For the second case in \eqref{eq:Jactioncone}, it is immediate that   $\phi(I_{\mathcal I_{\text{Big}}}\circ\sigma^{-1}\circ I_{\sigma(\mathcal I_{\text{Big}})}\circ I_{\sigma(\mathcal I)}\circ\sigma(x))\in\mathcal A_{\text{Big}}'\subset\mathcal A_{\mathcal I}'$. Wether $I_{\sigma(\mathcal I)}\circ\sigma(x) \in \sigma(\mathcal{I}_{\rm  Big})$ or $I_{\sigma(\mathcal I)}\circ\sigma(x) \in \sigma(\mathcal{I}_{\rm  Big})'$ is the same distinction found in \cite{AbateKoifman2023,Mintchev:2022fcp} by analytically continuing the modular evolution to $t=i\beta/2$.

We could use this to obtain expressions for the 2D case, considering different sign possibilities as in the previous subsection, and we would see the commutant in the 2D case is made from fields in the union of spacelike and timelike complements to $\mathcal{O}$, as in \cite{Mintchev:2022fcp}.

A nice feature of this purification is that {it allows one} to interpolate, at least heuristically,  between the modular conjugations of two different purifications of $\omega_\beta|_{\mathcal{A}_{\mathcal{O}}}$, one $J$ being close to the global thermal state $\omega_\beta$ modular conjugation and the other $J$ being the one associated to $\Omega|_{\mathcal{A}_{\mathcal{O}}}$. This can be seen, on one extreme, by enlarging $\mathcal{O}_{\text{Big}}$ to occupy the whole Minkowski space and this gives a purification of the thermal state without reducing to a local algebra and one approaches the same $J$ discussed in Section~\ref{globalpurifvacuumvector}. The opposite extreme would be to shrink $\mathcal{O}_{\text{Big}}$ all the way to $\mathcal{O}$, where now the purification corresponds to \eqref{tripletcone} but for the natural cone of $\mathcal{A}_{\mathcal{O}}$ and with some vector $|\Psi_{\mathcal{A}_{\mathcal{O}}}
\rangle$. Again, by property 7 of Proposition~\ref{prop:cone}, this immediately gives the modular conjugation $J_{|\Psi_{\mathcal{A}_{\mathcal{O}}}\rangle,\mathcal{O}}=J_{|\Omega\rangle,\mathcal{O}}$ of the vacuum state restricted to ${\mathcal{A}_{\mathcal{O}}}$ which acts by the inversion $I_{\mathcal{O}}$. What is happening is that as we enlarge or shrink $\mathcal{O}_{\text{Big}}$, different modular conjugations appear corresponding to different purifications of  $\omega_\beta|_{\mathcal{A}_{\mathcal{O}}}$, however all these modular conjugations always send $\mathcal{A}_{\mathcal{O}}$ to its commutant $\mathcal{A}_{\mathcal{O}}'$, but these antilinear maps $\mathcal{A}_{\mathcal{O}}\rightarrow \mathcal{A}_{\mathcal{O}}'$ are in principle all different.

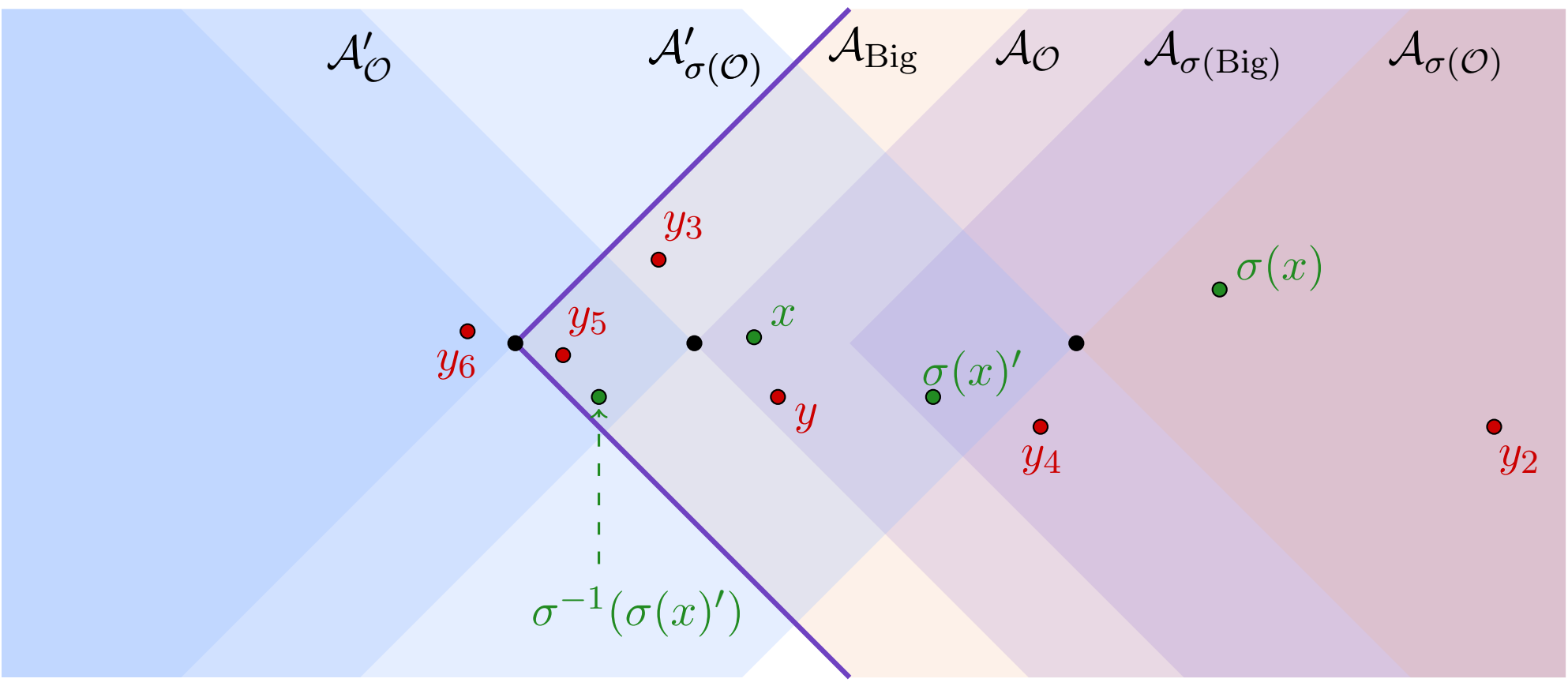
\begin{figure}[h!]
\centering
\begin{tikzpicture}[scale=1]


\def\a{1.5} 
\def\aa{2.8}
\def\b{4.7}
\def\Xmax{5.8}
\def\Xmin{2.8}
\def\Tmax{3}
\def\Tmaxx{4}


\begin{scope}
\clip
  (0,0)
  --
  (\Xmin,\Xmin)
  --
  (\Xmax+\Tmax,\Xmin)
  --
  (\Xmax+\Tmax,-\Xmin)
  --
  (\Xmin,-\Xmin)
  --
  cycle;

\fill[AsFill,opacity=.18]
  (-1,-\Xmax)
  rectangle
  (\Xmax+\Tmax,\Xmax);
\end{scope}


\begin{scope}
\clip
  (\a,0)
  --
 (\Xmin+\a,\Xmin)
  --
  (\Xmax+\Tmax,\Xmin)
  --
  (\Xmax+\Tmax,-\Xmin)
  --
  (\Xmin+\a,-\Xmin)
  --
  cycle;

\fill[ArFill,opacity=.18]
  (\a-1,-\Xmax)
  rectangle
  (\Xmax+\Tmax,\Xmax);
\end{scope}


\begin{scope}
\clip
  (\aa,0)
  --
 (\Xmin+\aa,\Xmin)
  --
  (\Xmax+\Tmax,\Xmin)
  --
  (\Xmax+\Tmax,-\Xmin)
  --
  (\Xmin+\aa,-\Xmin)
  --
  cycle;

\fill[ArFill,opacity=.18]
  (\aa-1,-\Xmax)
  rectangle
  (\Xmax+\Tmax,\Xmax);
\end{scope}


\begin{scope}
\clip
  (\b,0)
  --
  (\Xmin+\b,\Xmin)
  --
  (\Xmax+\Tmax,\Xmin)
  --
  (\Xmax+\Tmax,-\Xmin)
  --
  (\Xmin+\b,-\Xmin)
  --
  cycle;

\fill[AsFill,opacity=.18]
  (\b-1,-\Xmax)
  rectangle
  (\Xmax+\Tmax,\Xmax);
\end{scope}


\begin{scope}
\clip
  (\b,0)
  --
  (\b-\Xmin, \Xmin)
  --
  (-\Xmin-\Tmax, \Xmin)
  --  
  (-\Xmin-\Tmax,-\Xmin)
  --
  (\b-\Xmin, -\Xmin)
  --
  cycle;

\fill[AspFill,opacity=.23]
  (\Xmin+\b,\Xmin)
  rectangle
  (-\Xmin-\a,-\Xmin);
\end{scope}


\begin{scope}
\clip
  (\a,0)
  --
  (\a-\Xmin, \Xmin)
  --
  (-\Xmin-\Tmax, \Xmin)
  --  
  (-\Xmin-\Tmax,-\Xmin)
  --
  (\a-\Xmin, -\Xmin)
  --
  cycle;

\fill[AspFill,opacity=.23]
  (\Xmin+\b,\Xmin)
  rectangle
  (-\Xmin-\a,-\Xmin);
\end{scope}


\begin{scope}
\clip
  (0,0)
  --
  (0-\Xmin, \Xmin)
  --
  (-\Xmin-\Tmax, \Xmin)
  --  
  (-\Xmin-\Tmax,-\Xmin)
  --
  (0-\Xmin, -\Xmin)
  --
  cycle;

\fill[AspFill,opacity=.23]
  (\Xmin+\b,\Xmin)
  rectangle
  (-\Xmin-\a,-\Xmin);
\end{scope}






\draw[
  ArLine,
  line width=1.2pt
]
  (0,0)
  --
  (\Xmin,\Xmin);

\draw[
  ArLine,
  line width=1.2pt
]
  (0,0)
  --
  (\Xmin,-\Xmin);







\coordinate (x)   at (0.5+\a,0.05);
\coordinate (sx)  at (1.2+\b,0.45);
\coordinate (sxp) at (-1.2+\b,-0.45);
\coordinate (invsxp) at (-.8+\a,-0.45);


\filldraw[
  fill=ForestGreen,
  draw=black
]
  (x)
  circle (1.7pt);

\filldraw[
  fill=ForestGreen,
  draw=black
]
  (sx)
  circle (1.7pt);

\filldraw[
  fill=ForestGreen,
  draw=black
]
  (sxp)
  circle (1.7pt);

  \filldraw[
  fill=ForestGreen,
  draw=black
]
  (invsxp)
  circle (1.7pt);

\node[
  ForestGreen,
  anchor=west
]
  at ($(x)+(0,0.18)$)
  {$x$};

\node[
  ForestGreen,
  anchor=west
]
  at ($(sx)+(0,0.18)$)
  {$\sigma(x)$};

\node[
  ForestGreen,
  anchor=west
]
  at ($(sxp)-(0.23,-0.2)$)
  {$\sigma(x)'$};

  \node[
  ForestGreen,
  anchor=west
]
  at ($(invsxp)-(0.7,1.8)$)
  {$\sigma^{-1}(\sigma(x)')$};

\draw[
  ForestGreen,
  dash pattern=on 3pt off 4pt,
  opacity=1,
  line width=.6pt,
->]
  ($(invsxp)-(0,1.4)$)
  --
  ($(invsxp)-(0,0.1)$)
  ;


\coordinate (y)   at (2.2,-0.45);
\coordinate (sy)  at (\b+3.5,-.7);
\coordinate (syp) at (\b-3.5,.7);
\coordinate (y4) at (\aa+\aa-\b+3.5,-.7);
\coordinate (y5) at (.4,-.1);
\coordinate (y6) at (-.4,.1);
\coordinate (Is) at (\b,0);
\coordinate (Ia) at (\a,0);
\coordinate (Ibig) at (0,0);

\filldraw[
  fill=black!80!black,
  draw=black
]
  (Ia)
  circle (1.7pt);

\filldraw[
  fill=black!80!black,
  draw=black
]
  (Ibig)
  circle (1.7pt);

\filldraw[
  fill=black!80!black,
  draw=black
]
  (Is)
  circle (1.7pt);


\filldraw[
  fill=red!80!black,
  draw=black
]
  (y)
  circle (1.7pt);

\filldraw[
  fill=red!80!black,
  draw=black
]
  (sy)
  circle (1.7pt);

\filldraw[
  fill=red!80!black,
  draw=black
]
  (syp)
  circle (1.7pt);

\filldraw[
  fill=red!80!black,
  draw=black
]
  (y4)
  circle (1.7pt);

\filldraw[
  fill=red!80!black,
  draw=black
]
  (y5)
  circle (1.7pt);

 \filldraw[
  fill=red!80!black,
  draw=black
]
  (y6)
  circle (1.7pt);

\node[
  red!80!black,
  anchor=west
]
  at ($(y)+(0,-0.18)$)
  {$y$};

\node[
  red!80!black,
  anchor=west
]
  at ($(sy)+(-.1,-0.28)$)
  {$y_2$};

\node[
  red!80!black,
  anchor=west
]
  at ($(syp)+(-.1,0.28)$)
  {$y_3$};

\node[
  red!80!black,
  anchor=west
]
  at ($(y4)+(-.3,-0.28)$)
  {$y_4$};

\node[
  red!80!black,
  anchor=west
]
  at ($(y5)+(-.1,0.28)$)
  {$y_5$};  

\node[
  red!80!black,
  anchor=west
]
  at ($(y6)+(-.4,-0.28)$)
  {$y_6$};


\node at (3,2.45)
  {$\mathcal A_{\text{Big}}$};

\node at (\a+\Xmin,2.45)
  {$\mathcal A_{\mathcal O}$} ;

  \node at (\aa+\Xmin+.25,2.4)
  {$\mathcal A_{\sigma(\text{{Big}})}$} ;

\node at (\b+\Xmin+.3,2.4)
  {$\mathcal A_{\sigma(\mathcal O)}$};

\node at (\b-\Xmin-.3,2.4)
  {$\mathcal A'_{\sigma(\mathcal{O})}$};

\node at (\b-\Xmin-.3,2.4)
  {$\mathcal A'_{\sigma(\mathcal{O})}$};

  \node at (\a-\Xmin,2.4)
  {$\mathcal A'_{\mathcal{O}}$};

\end{tikzpicture}
\caption{The algebras associated to wedges relevant for the computation of the action of $J$ in the purification of the natural cone of some Big region $\mathcal A_\text{Big}$. We make the simplification of drawing wedges instead of double cones. The Big region has null boundaries in solid purple lines, and to their right we have the purification. We consider both cases in \eqref{eq:Jactioncone}: the first in green, starting from a field located at $x\in \mathcal{O}$ and calling $\sigma(x)'=I_{\sigma(\mathcal{O})}\circ \sigma(x)$. Then $\phi(\sigma(x)')$ belongs to  $\mathcal{A}_{\sigma(\text{Big})}$ by assumption and the final result is $\phi(\sigma^{-1}(\sigma(x)'))\in\mathcal{A}_{\text{Big}} \cap \mathcal{A}_{\mathcal{O}}'$, so it belongs to the purification. The second case is in red, and again starting from a field located at $y\in \mathcal{O}$. Then $\phi(y_3)$ belongs to the commutant of $\mathcal{A}_{\sigma(\text{Big})}$, namely its causal complement (for a double cone there is also the timelike complement). Now we make the following definitions: $y_2=\sigma(y)$,  $y_3=I_{\sigma(\mathcal{O})}(y_2)$, $y_4=I_{\sigma(\text{Big})}(y_3)$, $y_5=\sigma^{-1}(y_4)$, $y_6=I_{\text{Big}}(y_5)$. The final result of the conjugation by $J$ is $\phi(y_6)\in \mathcal{A}_{\mathcal{O}}'$ which does not belong to the purification.} 
\label{fig:natconepurification}
\end{figure}

These constructions also provide concrete realizations of the canonical
purification proposed by Sorce
\cite{sorce2025continuumcanonicalpurifications}. For a fixed algebra
$\mathcal A$ and thermal state $\omega_\beta|_{\mathcal A}$, let
$(\mathcal H,\pi,\Psi)$ be any of the cyclic and separating
realizations considered above, with modular conjugation $J$. Since
$a,b\in\mathcal A$ are bounded, Sorce's defining functional can be written as
\[
 \widehat\omega_\beta(a\otimes b^{\mathrm{op}})
 =\langle\pi(a^*)\Psi| J|\pi(b^*)\Psi\rangle
 =\langle\Psi|\pi(a)J\pi(b^*)J|\Psi\rangle .
\]
Thus the opposite algebra is represented by
$b^{\mathrm{op}}\mapsto J\pi(b^*)J$, and our computations identify
its action and localization in each realization. More precisely, the modular conjugation in the canonical purification of \cite{sorce2025continuumcanonicalpurifications} is given by $J_{\hat \omega} (a \otimes b)J_{\hat \omega}=b^*\otimes a^* $, whose localization structure in spacetime may not be as evident as in our purifications. However by the unitary equivalence
of all these representations, we can see how $J_{\hat\omega}$ works mapped to one of the purifications studied above.
We will give a more explicit connection to \cite{sorce2025continuumcanonicalpurifications} in the example that follows.

\section{A free massless scalar}\label{sec:scalar}

In addition to the purifications discussed in the previous section, we can have other purifications, in particular one where we essentially duplicate the Hilbert space, as is standard with finite degrees of freedom. This is what we are going to study in this Section for a massless scalar field, or more precisely, for the chiral U$(1)$ current.   

In this case, we have the canonical commutations relations which give rise to a Weyl $C^*$-algebra of Weyl operators. For a given Gaussian state, such as a thermal state, there is a canonical way to represent the Weyl algebra on a Fock space \cite{Petz:1990gb}.  As we will explain in detail, the construction uses a doubled real one-particle
space, equipped with a complex structure that mixes the two summands,
and its associated Fock space. Localization in $\mathcal I$ will enter
through the first real summand, via the vectors $f\oplus0$ with
$\operatorname{supp}f\subset\mathcal I$, which generate the local
algebra through their Weyl operators. However this will not mean that the
local algebra acts on only one Fock tensor factor.

Let us recall a few general facts about CCR algebras and quasi-free states. Given a real symplectic space $(\mathcal{Q},\sigma)$ we get a CCR algebra with relations
    \begin{equation}\label{CCR}
    W(f)W(g)=e^{-i\sigma(f,g)}W(f+g),\quad W(f)^*=W(-f)    
    \end{equation}
where $f,g \in \mathcal{Q}$. We will refer to this algebra as CCR$(\mathcal{Q},\sigma)$.

A positive symmetric bilinear form $\tau$ such that 
\begin{equation}\label{technical}
    \sigma(f,g)^2\leq \tau(f,f)\tau(g,g)
\end{equation}
defines a quasi-free state by \cite{Kay:1988mu} 
\begin{equation}
    \omega(W(f))=e^{-\frac{1}{2}\tau(f,f)}
\end{equation}
with 2-point function $w_2(f,g)=\tau(f,g)+i\sigma(f,g)$. The symplectic space $\mathcal{Q}$ can be completed using the inner product defined by $\tau$ and we shall call it $\mathcal{H}$. However the state $\omega$ may not be pure. It is pure if and only if $w_2$ is a complex inner product (as in the vacuum state). Let us assume $\langle\cdot,\cdot\rangle :=w_2(\cdot,\cdot)$ is a complex inner product for now (the complex structure will be mentioned later), and describe the Fock space $F(\mathcal{H})$ that serves as cyclic representation of this pure state, 
\begin{equation}
    \pi(W(f))=V(f)
\end{equation}  where $V(f)$:
    \begin{equation}\label{Vf}
        V(f)e^0=e^{-\frac{1}{2}\langle f,f\rangle}e^f,\qquad f \in \mathcal{H}. 
    \end{equation}
Here $e^f$ is an unnormalized coherent state,
\begin{equation}
    e^f:=1\oplus f\oplus\frac{1}{\sqrt{2!}} f \otimes f \oplus ...
\end{equation}
which satisfies,
\begin{equation}
    (e^f,e^g)=e^{\langle f,g\rangle}
\end{equation}
Calling $|\Omega\rangle:=e^0=1$ the vacuum vector
\begin{equation}
    (\Omega,V(f)\Omega)=e^{-\frac{1}{2}\langle f,f\rangle}=\omega(W(f))
\end{equation}
so we see that this is a representation of the state given by $\tau$.

We can define the local algebras associated to a given real-linear subspace $H\subset \mathcal{H}$:
\begin{equation}
    R(H):=\left\{V(f);\quad f\in H \right\}''
\end{equation}
Note that, due to the Weyl relations \eqref{CCR}, the commutant $R(H)'$ satisfies
\begin{equation}
    R(H)'=R(H')
\end{equation}
where $H'$ is the symplectic complement of $H$.

It turns out that $\Omega$ is cyclic (respectively separating) for $R(H)$ if and only if $H$ is cyclic (respectively separating). $H$ is cyclic if $\overline{H+iH}=\mathcal{H}$ while separating if $H\cap iH=0$. If $H$ is also  closed it is called a \textit{standard} subspace . 

Here we consider the derivative of the chiral part of the free massless scalar field in 1+1 dimensional Minkowski space.\footnote{The chiral current considered here transforms
under the vacuum modular conjugation as
$J_{\mathrm{Fock}}\phi(x)J_{\mathrm{Fock}}=-\phi(-x)$.
By antiunitarity, this is equivalent to
$J_{\mathrm{Fock}}V(f)J_{\mathrm{Fock}}=V(f\circ I_0)$
for $V(f)=e^{i\phi(f)}$. The additional sign does not affect the
localization arguments of Section~\ref{2dCFTs}.} This net is called $U(1)$-current net (see \cite{Camassa:2011wk}) and $\sigma$ is given by\footnote{Not to be confused with the conformal map of~Section \ref{2dCFTs}.}

\begin{equation}\label{eq:current-symplectic}
    \sigma(f,g)=\int_{\mathbb{R}}f(x) g'(x)dx.
\end{equation}  
It can be shown that $H=\bar{\left\{f\in\mathcal{C}_0^\infty(r,\infty)\right\}}$, namely the closure of the space of compactly supported real functions on some half-line $(r,\infty)$,  is a standard subspace \cite{Longo:2008zz}. The same holds for bounded intervals $\mathcal{I}\subset \mathbb{R}$. So we have a net of von Neumann algebras given by
\begin{equation}\label{localalgebraU(1)}
    \mathcal{A}_{\mathcal{I}} = \left\{V(f);\quad f\in\mathcal{C}_0^\infty(\mathcal{I})\right\}''
\end{equation}
with cyclic and separating vector $|\Omega\rangle$. Moreover, there is a relation between modular operators for each of these local algebras and modular operators associated to standard subspaces \cite{Longo:2008zz}. The Tomita operator $s_{\mathcal{I}}$ associated to $H(\mathcal{I})$ is defined by $s (f + i g)=f - i g$, where $f$ and $g$ belong to $H(\mathcal{I})$. Then one can take the polar decomposition of $s_{\mathcal{I}}$ and obtain the modular operator $\Delta_I$ and modular conjugation $J_{\mathcal{I}}$. The relation to the usual modular operators associated to $\mathcal{A}_{\mathcal{I}}$ is 
\begin{equation}
 \Delta_{\mathcal{A}_{\mathcal{I}},|\Omega\rangle}=e^{\Delta_{\mathcal{I}}}=1\oplus \Delta_{\mathcal{I}}
\oplus (\Delta_{\mathcal{I}} \otimes \Delta_{\mathcal{I}}) \oplus (\Delta_{\mathcal{I}} \otimes \Delta_\mathcal{I} \otimes \Delta_\mathcal{I})\oplus \cdots  
\end{equation}
With the convention \eqref{Vf}, the Fock modular conjugation is instead
\[
 J_{\mathcal{A}_{\mathcal{I}},|\Omega\rangle}
 =e^{-J_{\mathcal{I}}}
 =1\oplus(-J_{\mathcal{I}})\oplus(J_{\mathcal{I}}\otimes J_{\mathcal{I}})
 \oplus(-J_{\mathcal{I}}\otimes J_{\mathcal{I}}\otimes J_{\mathcal{I}})\oplus\cdots.
\]
Indeed, the Fock Tomita operator sends $e^f$ to $e^{-f}$ for $f\in H(\mathcal{I})$. This means that if we obtain the modular operators on the one-particle space, we have the corresponding operators of the von Neumann algebra acting on the Fock space.

Now, we have to consider the case when $\tau$ does not give a pure state, namely the two-point function $w_2$ is not a complex inner product. From \eqref{technical} it can be shown that a contraction $D$ exists such that 
\begin{equation}
    \sigma(f,g)=\tau(f,D g).
    \label{D}
\end{equation}
Because of the non-degeneracy of $\sigma$, $D$ has trivial kernel and $D=C|D|$ is its polar decomposition. $C$ is a complex structure and the state $\omega$ is pure if and only if $w_2$ is a complex inner product, bi-linear with respect to the complex structure $C$. This is actually equivalent to the statement that $\omega$ is pure if and only if $|D|=1$ (see \cite{Petz:1990gb} for further details).

The geometric thermal state $\tau_\beta$ is obtained from that of the vacuum,
\begin{equation}
    \tau_\beta(f,f):=\tau(f\circ\sigma^{-1},f\circ\sigma^{-1}),\qquad f\in \mathcal{C}_0^\infty(\mathbb{R})
\end{equation}
This is nothing but the same thermal state we have been considering before.
The one-particle \textit{real} Hilbert space $\mathcal{H}_\beta$ is obtained completing with respect to $\tau_\beta$,
\begin{equation}\label{thermaltwopointfunction}
    (f,g)_{\mathcal{H}_\beta}=\tau_\beta(f,g)=\operatorname{Re}w_2(f,g)=\operatorname{Re}(f\circ\sigma^{-1},g\circ\sigma^{-1})_\mathcal{H},\qquad f,\,g\,\in \mathcal{C}_0^\infty(\mathbb{R})
\end{equation}
which can be understood by noticing as before that precomposition with $\sigma^{-1}$ restricts the vacuum inner product to the positive real line. Of course the above expression can be checked explicitly, with the thermal two-point function given by
\begin{equation}
    w_2(f,g):=\langle \phi(f)\phi(g)\rangle_{\omega_\beta}=-\frac{1}{\pi}\int_{\mathbb{R}^2} dx\,dy\,\frac{f(x)g(y)}{\frac{\beta^2}{\pi^2}\sinh^2\left(\frac{\pi}{\beta}(x-y)-i\epsilon\right)}\label{analextension}
\end{equation}

\subsection{A purification of Gaussian state by doubling}

An arguably canonical purification of quasi-free states over a CCR-algebra \cite{ArakiShiraishi1971,Petz:1990gb} starts with 
\begin{equation}
    \mathcal{H}_\oplus=\mathcal{H}_\beta\oplus\mathcal{H}_\beta
\end{equation}
always at the one-particle level. The complex\footnote{\label{fn:doubled-complex-structure}A non-trivial point is that the complex structure ${i}_\oplus $ mixes both copies in $\mathcal{H}_\oplus$, this makes $\mathcal{H}_\oplus$ a complex Hilbert space such that the thermal state is a Fock state,  \cite{Petz:1990gb}. In other words, $i_\oplus$ is not diagonal as an operator on  $\mathcal{H}_\oplus$: $$i_\oplus=\left(\begin{matrix}
    -D & C\sqrt{1+D^2}\\
    C\sqrt{1+D^2} & D
\end{matrix}\right)$$.} inner product of $\mathcal{H}_\oplus$ is given by
    \begin{equation}
        \langle \cdot ,\cdot\rangle_\oplus=\tau_\oplus(\cdot,\cdot ) + i \sigma_\oplus(\cdot,\cdot ) ,
    \end{equation} 
with $\tau_\oplus:=\tau \oplus \tau$, where $\tau(f,f)=(f,f)_{\mathcal{H}_\beta}$ is a positive symmetric bilinear form, and $\sigma_\oplus(\cdot,\cdot)=\tau_\oplus(\cdot,-i_{\oplus} \cdot)$. This inner product  reduces to the thermal two-point function on $\mathcal{H}_\beta\simeq \mathcal{H}_\beta\oplus 0$:
\begin{equation}
    \langle f\oplus 0,g\oplus 0\rangle_\oplus=\tau_\beta(f,g)+i\sigma(f,g)=w_2(f,g)
    \end{equation}

The Fock space $F(\mathcal{H}_\oplus)=F(\mathcal{H}_\beta\oplus\mathcal{H}_\beta)$ is isomorphic to the product of Fock spaces $F(\mathcal{H}_\beta)\otimes F(\mathcal{H}_\beta)$. The representation of the quasi-local algebra is:
\begin{equation}\label{pibetafock}
    \pi_{\beta\oplus}(V(f))=V_\oplus(f\oplus 0), 
\end{equation}
such that, 
\begin{equation}\label{Vfplus}
    V_\oplus(f \oplus g) e^{0\oplus 0} =e^{-\frac{1}{2}||f\oplus g||_{\mathcal{H}_\oplus}^2} e^{f\oplus g}
\end{equation}
It is then straightforward to show that \eqref{pibetafock} defines a cyclic representation on a Fock space associated to the thermal state $\omega_\beta$, with cyclic vector $|\Omega_\oplus\rangle:=e^{0\oplus 0}\simeq 1 \otimes 1$:
\begin{equation}
    \left( |\Omega_\oplus\rangle,\pi_{\beta\oplus}(V(f))  |\Omega_\oplus\rangle\right) =\left( e^{0\oplus 0}, e^{-\frac{1}{2}||f\oplus 0||_{\mathcal{H}_\oplus}^2} e^{f\oplus 0}\right) = e^{-\frac{1}{2}||f||^2_{\mathcal{H}_\beta}}=\omega_\beta(V(f))
\end{equation}

The real subspaces $H(\mathcal{I})\simeq H_\oplus(\mathcal{I})=\bar{\left\{f\oplus 0;\quad f\in\mathcal{C}_0^\infty(\mathcal{I})\right\}}$ made of functions $f$ with support on $\mathcal{I}$ are cyclic and separating with respect to $\mathcal{H}_\oplus$ \cite{Garbarz:2021rqu}, so the vector $|\Omega_\oplus\rangle$ is cyclic and separating for the representation \eqref{pibetafock} of the local algebras \eqref{localalgebraU(1)}.

\subsection{Modular operators from the vacuum representation}

Let us show that the modular evolution $\Delta_{\oplus,\mathbb{R}_+}^{is}$ on the one-particle Hilbert space $\mathcal{H}_\oplus$ can be obtained easily with the aid of the unitary\footnote{It is an isometric map between standard subspaces: $H({\mathbb{R}_+})\rightarrow H(\mathbb{R})\oplus 0$, which by complex linearity extends to a map between dense subspaces, so it can be extended to a unique unitary operator.},
\begin{equation}\label{u}
    u (f \circ \sigma^{-1}) := f \oplus 0, \qquad f\in\mathcal{C}_0^\infty(\mathbb{R})
\end{equation}
This unitary comes from the unitary $U_\oplus:  F(\mathcal{H})\rightarrow F(\mathcal{H}_\oplus)$ intertwining the cyclic representations 
\begin{equation}
    U_\oplus \pi_\beta(V(f))\,|\Omega\rangle=\pi_{\beta\oplus}(V(f))\,|\Omega_\oplus\rangle
\end{equation}
which using \eqref{pibetaglobal}, \eqref{Vf},  \eqref{thermaltwopointfunction}, \eqref{pibetafock} and \eqref{Vfplus} reads
\begin{equation}
    U_\oplus e^{f\circ\sigma^{-1}}=e^{f\oplus 0} 
\end{equation}
In essence, this says that we can trade the geometric way of representing/defining the thermal state with adding a copy of Fock space $F(\mathcal{H})$. 

By the same reasoning of the previous section using covariance and the Bisognano-Wichmann theorem\footnote{There is an analogous statement for standard subspaces \cite{Longo:2008zz}. In short, the modular evolution associated to the positive real line is given by dilations: $\Delta_{\mathbb{R}_+}^{is}f(x)=f(e^{2\pi s}x)$. While the modular conjugation reads $J_{\mathbb{R}_+}f(x)=-f(-x)$. } 
we can assert,
\begin{equation}
    \Delta^{is}_{\oplus,\mathbb{R}_+}=u \Delta_{\sigma(\mathbb{R}_+)}^{is} u^{-1} 
\end{equation}
It is useful for what follows to see explicitly how $\Delta^{is}_{\oplus,\mathbb{R}_+}$ acts on functions supported in the positive real line:
\begin{align}
    \left[\left(u\, \Delta_{\sigma(\mathbb{R}_+)}^{is}\, u^{\dagger} \right) f\oplus 0\right](x)&= u\,f\left[\frac{\beta}{2\pi}\log\left(1+e^{2\pi s}\left(x-1\right)\right)\right]
\end{align}
The function $f\left[\frac{\beta}{2\pi}\log\left(1+e^{2\pi s}\left(x-1\right)\right)\right]$ is supported in $\sigma(\mathbb{R}_+)$  so we know how to act on it with $u$ using \eqref{u}. Therefore,
\begin{equation}
    \Delta_{\oplus,\mathbb{R}_+}^{is}f(x)\oplus 0=f\left[\frac{\beta}{2\pi}\log\left(1+e^{2\pi s}\left(e^{\frac{2\pi x}{\beta}}-1\right)\right)\right] \oplus 0
\end{equation}
which agrees with the thermal modular flow of a half-line found in \textcolor{blue}{\cite[Theorem~4.2]{BorchersYngvason1999}}.

We now turn to the one-particle modular conjugations
\(J_{\oplus,\mathbb R_+}\) and \(J_{\oplus,\mathbb R}\), associated with
\(H_\oplus(\mathbb R_+)\) and
\(H_\oplus(\mathbb R)=\mathcal H_\beta\oplus0\), respectively. The discussion following \eqref{Jhalfline} already identifies the interval
\((r,\sigma^{-1}(2\sigma(r)))\) whose reflected image stays in the
original thermal representation. Since
\(\sigma^{-1}(2\sigma(r))=r+\frac{\beta}{2\pi}\log2\), its endpoint
for \(r=0\) is \(\sigma^{-1}(2)=\frac{\beta}{2\pi}\log2\).
We will now recover that same distinction in the doubled representation.

For real \(f\in \mathcal{C}_0^\infty((0,\infty))\), the vacuum standard-subspace conjugation acts on current smearings as
\(J_{\mathbb R_+}f(x)=-f(-x)=-f \circ I_0 (x)\)%
\footnote{With \(\widehat f(p)=\int e^{ipx}f(x)\,dx\), the vacuum
modular dilations give
\(\widehat{\Delta_{\mathbb R_+}^{it}f}(p)
=e^{-2\pi t}\widehat f(e^{-2\pi t}p)\), \(p>0\).
For real \(f\in \mathcal{C}_0^\infty((0,\infty))\), the boundary value at
\(t=-i/2\) is \(-\widehat f(-p)\).
Since \(J_{\mathbb R_+}f=\Delta_{\mathbb R_+}^{1/2}f\) on
\(H(\mathbb R_+)\), the minus sign follows.}.
With the same reasoning as above for $\Delta^{it}_{\oplus,\mathbb{R}_+}$, we have
\begin{equation}
 \left(J_{\oplus,\mathbb R_+}(f\oplus0)\right)
 =uJ_{\sigma(\mathbb R_+)}(f\circ\sigma^{-1})
 =-u\bigl[f\circ\sigma^{-1}\circ I_{\sigma(0)}\bigr].
 \label{eq:sw42-transport}
\end{equation}
If \(\operatorname{supp}f\subset(0,\sigma^{-1}(2))\), the reflected
function $f\circ\sigma^{-1}\circ I_{\sigma(0)}$ is supported in \((0,1)\), so \eqref{u} applies directly:
\begin{equation}
 \bigl[J_{\oplus,\mathbb R_+}(f\oplus0)\bigr](x)
 =-f\!\left(\frac{\beta}{2\pi}
       \log\bigl(2-e^{2\pi x/\beta}\bigr)\right)\oplus0.
 \label{Jplus}
\end{equation}
The function on the right is understood to vanish outside its real
branch; its support is a compact subset of \(( -\infty,0)\).
Thus vectors $f \oplus 0$ in \(H_\oplus(\mathbb R_+)\), with \(\operatorname{supp}f\subset(0,\sigma^{-1}(2))\),  have modular images
entirely in the first real summand.

For the other case where \(\operatorname{supp}f\subset(\sigma^{-1}(2),\infty)\), we can do a similar procedure as we did in the previous section: use the fact that 
\(uJ_{\mathbb R_+}u^*=J_{\oplus,\mathbb R}\) as well as $-J_{\mathbb R_+}f \circ I_0=f$, 
\begin{equation}
 uf \circ \sigma^{-1}\circ  I_{\sigma(0)} =-\,u  J_{\mathbb R_+}f \circ \sigma^{-1}\circ  I_{\sigma(0)}\circ I_0=-J_{\oplus,\mathbb R}\,u f \circ \sigma^{-1}\circ  I_{\sigma(0)}\circ I_0
 \label{eq:sw42-negative}
\end{equation}
Consequently, if \(\operatorname{supp}f\subset(\sigma^{-1}(2),\infty)\),
\begin{equation}
 J_{\oplus,\mathbb R_+}(f\oplus0)
 =J_{\oplus,\mathbb R}(h\oplus0),\qquad
 h(x)=f\!\left(\frac{\beta}{2\pi}
                   \log\bigl(2+e^{2\pi x/\beta}\bigr)\right).
 \label{eq:sw42-far}
\end{equation}
Here \(h\) is real and compactly supported. This image belongs to
\(H_\oplus(\mathbb R)'\), the global symplectic complement. We now want to compute the global conjugation in the doubled representation $J_{\oplus,\mathbb R}$ and we will see it mixes the two components.

We compute \(J_{\oplus,\mathbb R}\) directly from the observation that it should map to the symplectic complement, which in Fock space gives the commutant, see \eqref{CCR}. Our Fourier and translation
conventions are
\begin{equation}
 \widehat f(p)=\int_{\mathbb R}e^{ipx}f(x)\,dx,\qquad
 T(t)f(x)=f(x-t),\qquad
 \widehat{T(t)f}(p)=e^{ipt}\widehat f(p).
 \label{eq:sw42-fourier}
\end{equation}
For the geometric current state, the real thermal covariance is \cite[Theorem~4.3]{Camassa:2011wk}
\begin{equation}
 \tau_\beta(f,g)=\frac1\pi\operatorname{Re}\int_0^\infty
 p\coth\!\left(\frac{\beta p}{2}\right)
 \overline{\widehat f(p)}\widehat g(p)\,dp.
 \label{eq:sw42-covariance}
\end{equation}
Together with the symplectic form \eqref{eq:current-symplectic}, this gives, for real \(f,g\),
\begin{equation}
 w_2(f,g)=\frac1\pi\int_0^\infty dp\,p\,
 \frac{\overline{\widehat f(p)}\widehat g(p)
       +e^{-\beta p}\widehat f(p)\overline{\widehat g(p)}}{1-e^{-\beta p}}.
 \label{eq:sw42-two-point}
\end{equation}
From \(\tau_\beta(f,Dg)=\sigma(f,g)\), we obtain
\begin{equation}
 \widehat{|D|f}(p)=\tanh\!\left(\frac{\beta p}{2}\right)\widehat f(p),
 \qquad
 \widehat{\sqrt{1+D^2}\,f}(p)
 =\frac{\widehat f(p)}{\cosh(\beta p/2)},\qquad p>0.
 \label{eq:sw42-multipliers}
\end{equation}
Both positive operators have trivial kernel, commute, and satisfy
\(|D|^2+(\sqrt{1+D^2})^2=1\).

With \(D=C|D|\), the complex structure in footnote~\ref{fn:doubled-complex-structure} gives
\(\sigma_\oplus(f\oplus0,g\oplus h)
=\tau_\beta(f,C(|D|g-\sqrt{1+D^2}\,h))\). Hence
\begin{equation}
 \begin{split}
 H_\oplus(\mathbb R)'
 &=\{g\oplus h:|D|g=\sqrt{1+D^2}\,h\}\\
 &=\{\sqrt{1+D^2}\,h\oplus|D|h:h\in\mathcal H_\beta\}.
 \end{split}
 \label{eq:sw42-complement}
\end{equation}
Notice that the map
\(h\mapsto\sqrt{1+D^2}\,h\oplus|D|h\) is a real isometry onto this symplectic
complement. Since \(J_{\oplus,\mathbb R}\) maps
\(H_\oplus(\mathbb R)\) onto \(H_\oplus(\mathbb R)'\), there is a
real orthogonal operator \(U\) on \(\mathcal H_\beta\) such that
\(J_{\oplus,\mathbb R}(f\oplus0)
=\sqrt{1+D^2}\,Uf\oplus|D|Uf\).

The Tomita identity
\(S_{\oplus,\mathbb R}(f\oplus0)=f\oplus0\), together with
\(S_{\oplus,\mathbb R}=J_{\oplus,\mathbb R}\Delta_{\oplus,\mathbb R}^{1/2}\),
implies
\begin{equation}
 \begin{split}
 \tau_\beta(g,\sqrt{1+D^2}\,Uf)
 &=\langle g\oplus0,J_{\oplus,\mathbb R}(f\oplus0)\rangle_\oplus\\
 &=\langle g\oplus0,\Delta_{\oplus,\mathbb R}^{1/2}(f\oplus0)\rangle_\oplus.
 \end{split}
 \label{eq:sw42-positivity}
\end{equation}
Then positivity of
$\Delta_{\oplus,\mathbb R}^{1/2}$ implies that the bounded
operator $\sqrt{1+D^2}\,U$ is positive and self-adjoint
with respect to $\tau_\beta$. So its square equals its product with its adjoint,
which is \(1+D^2\), and uniqueness of the positive square root gives
\(\sqrt{1+D^2}\,U=\sqrt{1+D^2}\). Furthermore, injectivity of  \(\sqrt{1+D^2}\) gives \(U=1\).
Therefore, taking into account \(J_{\oplus,\mathbb R}\) should be anti-linear with respect to \(i_\oplus\), the only possibility is,
\begin{equation}
 J_{\oplus,\mathbb R}=
 \begin{pmatrix}
  \sqrt{1+D^2}&|D|\\
  |D|&-\sqrt{1+D^2}
 \end{pmatrix}.
 \label{eq:sw42-conjugation}
\end{equation}
As anticipated, it sends \(f\oplus0\) to
\(\sqrt{1+D^2}\,f\oplus|D|f\), whose second component is nonzero
for \(f\ne0\). 

As a check, let \(w_2(f,T(i\beta/2)g)\) denote the scalar analytic
continuation of \(t\mapsto w_2(f,T(t)g)\) to the midpoint of the KMS
strip. Since under \(g\mapsto T(t)g\), the two terms in \eqref{eq:sw42-two-point} acquire the opposite
phases \(e^{ipt}\) and \(e^{-ipt}\), equation~\eqref{eq:sw42-two-point} gives
\begin{equation}
 \begin{split}
 w_2(f,T(i\beta/2)g)
 &=\frac1\pi\int_0^\infty
 \frac{p\,\operatorname{Re}(\overline{\widehat f(p)}\widehat g(p))}
      {\sinh(\beta p/2)}\,dp\\
 &=\tau_\beta(f,\sqrt{1+D^2}\,g)
 =\langle f\oplus0,J_{\oplus,\mathbb R}(g\oplus0)\rangle_\oplus.
 \end{split}
 \label{eq:sw42-half-strip}
\end{equation}
where we used the first line in \eqref{eq:sw42-positivity}. Equivalently,
\begin{equation}
 w_2(f,T(i\beta/2)g)=\frac{\pi}{\beta^2}
 \int_{\mathbb R^2}
 \frac{f(x)g(y)}{\cosh^2(\pi(x-y)/\beta)}\,dx\,dy.
 \label{eq:sw42-kernel}
\end{equation}
gives the matrix elements of $J_{\oplus,\mathbb R}$ in coordinate space between vectors in \(H_\oplus(\mathbb{R})\) and by anti-linearity determines all its matrix elements. 

In summary, for real $f\in\mathcal C_0^\infty(\mathbb R_+)$, define
\[
 f_-(x)=f\!\left(\frac{\beta}{2\pi}
       \log(2-e^{2\pi x/\beta})\right),\qquad
 h(x)=f\!\left(\frac{\beta}{2\pi}
       \log(2+e^{2\pi x/\beta})\right).
\]
Here $f_-$ is extended by zero where the logarithm has no real
argument. Combining \eqref{Jplus}, \eqref{eq:sw42-far} and
\eqref{eq:sw42-conjugation}, we obtain
\begin{equation}
 J_{\oplus,\mathbb R_+}(f\oplus0)=
 \begin{cases}
  -f_-\oplus0,
    &\operatorname{supp}f\subset(0,\sigma^{-1}(2)),\\[2pt]
  \sqrt{1+D^2}\,h\oplus|D|h,
    &\operatorname{supp}f\subset(\sigma^{-1}(2),\infty).
 \end{cases}
 \label{eq:sw42-halfline-summary}
\end{equation}
In the first
case $f_-$ is compactly supported on the negative half-line, while 
in the second case $h$ is compactly supported on the whole real line,
and the image lies in $H_\oplus(\mathbb R)'$. For these reasons, in both cases the image lies in $H_\oplus(\mathbb R_+)'$. If the support of $f$
crosses $\sigma^{-1}(2)$, the general formula
\eqref{eq:sw42-transport} still applies, and  one should not obtain this
case by multiplying $f$ by sharp characteristic functions since the
resulting pieces need not belong separately to the one-particle Hilbert space.

For $\mathcal I=(a,b)$, the same unitary $u$ maps
$H(\sigma(\mathcal I))$ onto $H_\oplus(\mathcal I)$. Hence
\begin{equation}
 J_{\oplus,\mathcal I}=uJ_{\sigma(\mathcal I)}u^*,\qquad
 \Delta_{\oplus,\mathcal I}^{it}
 =u\Delta_{\sigma(\mathcal I)}^{it}u^*.
 \label{eq:sw42-interval-transport}
\end{equation}
Thus the half-line reflection $I_{\sigma(0)}$ is replaced by the
interval inversion used in Section~\ref{sec:positivecone}, namely\footnote{ The inversion acts on
the compactified real line, so that at its pole it should be interpreted
through vacuum M\"obius covariance.
}
\begin{equation}
 I_{\sigma(\mathcal I)}(y)
 =\frac{\frac{\sigma(a)+\sigma(b)}{2}\,y-\sigma(a)\sigma(b)}
        {y-\frac{\sigma(a)+\sigma(b)}{2}}.
 \label{eq:sw42-interval-inversion}
\end{equation}
For real $f\in\mathcal C_0^\infty(\mathcal I)$ this gives
\begin{equation}
 J_{\oplus,\mathcal I}(f\oplus0)
 =-u\bigl[(f\circ\sigma^{-1})\circ I_{\sigma(\mathcal I)}\bigr].
 \label{eq:sw42-interval-action}
\end{equation}
To make the two support cases explicit, put
$g=(f\circ\sigma^{-1})\circ I_{\sigma(\mathcal I)}$.
If $g$ has compact support contained in the positive or negative
vacuum half-line, respectively, the same argument as before gives
\begin{equation}
 J_{\oplus,\mathcal I}(f\oplus0)=
 \begin{cases}
  -(g\circ\sigma)\oplus0,
       &\operatorname{supp}g\subset(0,\infty),\\[2pt]
  \sqrt{1+D^2}\,h\oplus|D|h,\quad h=g\circ I_0\circ\sigma,
       &\operatorname{supp}g\subset(-\infty,0).
 \end{cases}
 \label{eq:sw42-interval-cases}
\end{equation}
In particular, the second case occurs when the compact support of
$f$ is contained in
\[
 \left(
 \sigma^{-1}\!\left(\frac{2\sigma(a)\sigma(b)}{\sigma(a)+\sigma(b)}\right),
 \sigma^{-1}\!\left(\frac{\sigma(a)+\sigma(b)}{2}\right)
 \right),
\]
since this is precisely where $I_{\sigma(\mathcal I)}(\sigma(x))<0$.
Outside this open interval, the reflected points are positive,
except at the two endpoints, which map to zero and infinity.
For supports crossing either endpoint, or whose reflected image
meets both half-lines, \eqref{eq:sw42-interval-action} remains the
general formula.

\subsection{Comparison to Sorce's continuum purification}\label{sec:sorce-comparison}

In \cite{sorce2025continuumcanonicalpurifications}, it is shown that a way to purify a state $\omega$ defined on a *-algebra $\mathcal{A}$ is by first doubling as $\mathcal{A}\otimes \mathcal{A}^{\rm op}$, with  $\mathcal{A}^{\rm op}$ having the opposite-order algebra product, and using the GNS representation of $\omega$ define the purifying state
\begin{equation}
    \hat{\omega}(a \otimes b)=\langle a^*\Omega|J_\omega|b^*\Omega\rangle
\end{equation}
In the case $\mathcal{A}$ is a von Neumann algebra of bounded operators, domain issues are absent and we can go further and write
\begin{equation}
    \hat{\omega}(a \otimes b)=\langle \Omega|aJ_\omega b^*J_\omega|\Omega\rangle
    \label{eq:sorce-purifyingstate}
\end{equation}
(in this subsection $|\Omega\rangle$ denotes the GNS vector of $\omega$, not necessarily the vacuum). Assuming also that $\omega$ is faithful, its GNS vector is separating and we have a unitary that intertwines the GNS of $\omega$ with that of the GNS of $\hat{\omega}$, given by
\begin{equation}
    a\otimes b|\hat\Omega\rangle \mapsto a J_\omega b^* J_\omega |\Omega\rangle
\end{equation}

For the quasi-free states considered in this section, this canonical purification $\hat\omega$ coincides with the doubled Fock purification \eqref{pibetafock}. Indeed, it is shown in \cite[Section~5.1]{Caminiti:2026excitability} that the canonical purification of a Gaussian state is again Gaussian. In our notation, it is the quasi-free state on the Weyl algebra of the doubled space of pairs $f\oplus g$ of real test functions, with symplectic form $\sigma\oplus(-\sigma)$, so that the two summands commute and the second one generates the copy $\mathcal{A}^{\rm op}$, with covariance (up to normalization conventions, and with our $D$ playing the role of the operator $R$ of \cite{Caminiti:2026excitability})
\begin{equation}
 \hat\tau(f\oplus g,f\oplus g)=\tau_\beta(f,f)+\tau_\beta(g,g)+2\,\tau_\beta\bigl(f,\sqrt{1+D^2}\,g\bigr).
 \label{eq:sorce-covariance}
\end{equation}
As noted in \cite{Caminiti:2026excitability}, this is the formula of \cite[Eq.~(3.9)]{ArakiYamagami1982} for the doubling of a quasi-free state. It is the same state as ours, written in a different basis of the doubled one-particle space. In order to see this, consider the real-linear map
\begin{equation}
 \Phi(f\oplus g):=f\oplus0+J_{\oplus,\mathbb R}(g\oplus0)
 =\bigl(f+\sqrt{1+D^2}\,g\bigr)\oplus|D|g,
 \label{eq:sorce-map}
\end{equation}
where we used \eqref{eq:sw42-conjugation}. Since $|D|$ and $\sqrt{1+D^2}$ are self-adjoint with respect to $\tau_\beta$, commute, and satisfy $|D|^2+(\sqrt{1+D^2})^2=1$, we get
\begin{equation}
 \|\Phi(f\oplus g)\|^2_{\mathcal{H}_\oplus}=\tau_\oplus(
 \Phi(f\oplus g),\Phi(f\oplus g))
=\tau_\beta(f,f)+\tau_\beta(g,g)+2\,\tau_\beta\bigl(f,\sqrt{1+D^2}\,g\bigr),
\end{equation}
which is \eqref{eq:sorce-covariance}.  This means $\hat\tau$ and $\tau_\oplus$ define the same state. Moreover, $\Phi$ maps the second summand onto the symplectic complement $H_\oplus(\mathbb R)'$ described in \eqref{eq:sw42-complement}, and, $J_{\oplus,\mathbb R}$ being antiunitary, $\sigma_\oplus(\Phi(0\oplus g),\Phi(0\oplus h))=-\sigma(g,h)$, so $\Phi$ also intertwines the symplectic forms. Therefore the canonical purification of $\omega_\beta$ is realized on $F(\mathcal{H}_\oplus)$ by the vector $|\Omega_\oplus\rangle$ and the representation
\begin{align}
 \hat\pi\bigl(V(f)\otimes V(g)\bigr)
 &=\pi_{\beta\oplus}(V(f))\,J_{\mathrm{Fock}}\,\pi_{\beta\oplus}(V(g))^*J_{\mathrm{Fock}}\nonumber\\ &=V_\oplus(f\oplus0)\,V_\oplus\bigl(J_{\oplus,\mathbb R}(g\oplus0)\bigr)\nonumber\\
 &=V_\oplus\bigl(\Phi(f\oplus g)\bigr),
\end{align}
Here $J_{\mathrm{Fock}}=e^{-J_{\oplus,\mathbb R}}$ denotes the modular conjugation of the global algebra in the doubled Fock representation, which satisfies $J_{\mathrm{Fock}}V_\oplus(h)J_{\mathrm{Fock}}=V_\oplus(-J_{\oplus,\mathbb R}h)$; this gives the second equality, and the last equality holds because $f\oplus0$ and $J_{\oplus,\mathbb R}(g\oplus0)$ have vanishing symplectic form.

In the variables $f\oplus g$ the global modular conjugation is simply the swap, $J_{\oplus,\mathbb R}\Phi(f\oplus g)=\Phi(g\oplus f)$, in agreement with \cite{sorce2025continuumcanonicalpurifications}. While in our variables, where the real inner product $\tau_\beta\oplus\tau_\beta$ is diagonal, it takes the form \eqref{eq:sw42-conjugation}. Finally, the cross term in \eqref{eq:sorce-covariance}, which carries the correlations between the two copies, is precisely the half-strip continuation \eqref{eq:sw42-half-strip} of the thermal two-point function, with kernel \eqref{eq:sw42-kernel}.

\subsection{Other thermal states}

The $U(1)$ chiral net admits a family of primary KMS states $\omega_q$ parameterized by a real number $q$, as explained in detail in \cite{Camassa:2011wk}. Here primary means that the GNS representation is factorial, and the states $\omega_q$ exhaust the primary locally normal KMS states of the net \cite[Theorem~4.3]{Camassa:2011wk}. Since the canonical purification of a state is pure if and only if the state is factorial \cite[Section~3.5]{sorce2025continuumcanonicalpurifications}, the $\omega_q$ are exactly the KMS states of the net whose canonical purification is pure.

It is easy to motivate the existence of these $\omega_q$ states, by noticing that the Fourier decomposition of the chiral current $\partial\phi$ has a zero mode, the momentum, which commutes with every other mode. Then, a shift $\partial\phi+q$ preserves the algebra. More precisely, let $\gamma_q$ be the automorphism of the quasi-local algebra $\mathcal{A}_{\mathbb{R}}$,
\begin{equation}
    \gamma_q(V(f)):=e^{iq\int_\mathbb{R}f(x)dx}V(f).
\end{equation}
It can be locally unitarily implementable: for $V(f)\in \mathcal{A}_\mathcal{I}$, pick some compactly supported function $s_\mathcal{I}$ such that is the identity in $\mathcal{I}$, then\footnote{We have a factor $\frac12$ compared to \cite{Camassa:2011wk} since our definition of the CCR algebra \eqref{CCR} does not have a $\frac12$ in front of $\sigma$.},
\begin{equation}
    \gamma_q(V(f))=\text{Ad}\left(V\left(\frac{q}{2}s_\mathcal{I}  \right)\right)V(f), \qquad \text{supp}\,f
\subset \mathcal{I}.\end{equation}
We can set 
\begin{equation}
    \gamma_q|_\mathcal{I}:=\text{Ad}\left(V\left(\frac{q}{2}s_\mathcal{I}  \right)\right), \qquad U_{q,\mathcal{I}}:=V\left(\frac{q}{2}s_\mathcal{I}  \right).
\end{equation}
The family of KMS states $\omega_q$ is then defined as
\begin{equation}
    \omega_q:=\omega_\beta \circ \gamma_q.
\end{equation}
The case $q=0$ coincides with the geometric KMS state, $\omega_\beta=\omega_q|_{q=0}$. Taking into account the previous lines, we have the restriction $\omega_q|_{\mathcal{I}}=\omega_\beta|_\mathcal{I} \circ \gamma_q|_\mathcal{I}$. We can represent $\omega_q|_{\mathcal{I}}$ with,
\begin{equation} \label{qpurification}
    (\mathcal{H}\,,\quad\pi_q|_\mathcal{I}:=\pi_\beta|_\mathcal{I} \circ \gamma_q|_\mathcal{I}\,,\quad |\Omega\rangle).
\end{equation}
Indeed,
\begin{equation}
    \langle \Omega|{\rm Exp} \circ \gamma_q \,V(f)|\Omega\rangle=\omega_\beta\left(\gamma_q(V(f)) \right)=\omega_\beta\circ\gamma_q\,V(f)=\omega_q(V(f)).
\end{equation}

Now, the modular operators in the purification \eqref{qpurification} are immediate to obtain from the ones of the geometric KMS state, since the representation of $\mathcal{A}_\mathcal{I}$ is in $\mathcal{A}_\mathcal{\sigma(I)}$ with vector representative $|\Omega\rangle$, exactly as with the representation $\pi_\beta$ of the geometric KMS state . As we discussed already, the modular operators only depend on the algebra and cyclic and separating vector, so to  $\pi_q|_{\mathcal{I}}$   corresponds the modular conjugation $J_{q,\mathcal{I}}$ given by 
$J_{\beta,\mathcal{I}}$:
\begin{equation}
    J_{q,\mathcal{I}}=J_{\beta,\mathcal{I}}=J_{|\Omega\rangle,\sigma(\mathcal{I})},
\end{equation}
This means,
\begin{equation}
    J_{q,\mathcal{I}}\,\,\pi_q(V(f))\,\,
    J_{q,\mathcal{I}}=e^{-iq\int_\mathcal{I} f }J_{\beta,\mathcal{I}}\,\,\pi_\beta(V(f))\,\,
    J_{\beta,\mathcal{I}}.
\end{equation}



\section{Conclusions}

We have computed the modular conjugation $J$ of the geometric KMS state of a two-dimensional CFT in several explicit cyclic purifications, and in each of them we have followed where the operators $J\pi(a)J$ are localized.

In the purification with the vacuum vector, the thermal representation is the conformal map ${\rm Exp}$ onto the algebra of the right Rindler wedge (the positive half-line in a chiral component), and $J$ follows from the Bisognano-Wichmann theorem and covariance. For the global algebra the commutant lies entirely outside the image of the representation, see \eqref{Jglobal}. For the algebra of a wedge (a half-line $(r,\infty)$ in a chiral component) the commutant is larger and part of it lies in the image of the representation: the fields localized in $(r,\sigma^{-1}(2\sigma(r)))$ are mapped to thermal fields on the complementary half-line, while the remaining ones are mapped outside the image, see \eqref{Jhalfline}. In two dimensions the latter operators are localized not only in the left wedge but also in the future and past light cones.

In the purification with a vector in the natural cone of the algebra $\mathcal{A}_{\text{Big}}$ of a larger region $\mathcal{O}_{\text{Big}}\supseteq\mathcal{O}$ (Section~\ref{sec:positivecone}), the representation is the identity and the action of $J$ is given by \eqref{eq:Jactioncone}. Enlarging or shrinking $\mathcal{O}_{\text{Big}}$ allows to move between the previous purification and the one in the natural cone of $\mathcal{A}_{\mathcal{O}}$ itself, for which $J$ is the vacuum modular conjugation of $\mathcal{O}$. All these modular conjugations map $\mathcal{A}_{\mathcal{O}}$ onto its commutant, but they do so in different ways.

In all these cases the modular images are fields at real points. A ``second world'', or a field at a complex point $x\pm i\beta/2$, appears only if one insists that the whole commutant be in the image of the thermal representation, that is, if one tries to undo the map $\sigma$ on a point outside its image.

For the chiral $U(1)$ current we also considered the purification on the Fock space of a doubled one-particle space, the analogue of the usual purification for finite degrees of freedom. There the global modular conjugation \eqref{eq:sw42-conjugation} mixes the two copies, and for half-lines and intervals the modular images either stay in the first copy or belong to the global commutant, depending on the support of the smearing function, see \eqref{eq:sw42-halfline-summary} and \eqref{eq:sw42-interval-cases}. We showed that this purification coincides with the canonical purification of \cite{sorce2025continuumcanonicalpurifications}, in the Gaussian form of \cite{Caminiti:2026excitability}, through the explicit map \eqref{eq:sorce-map}, and that the correlations between the two copies are given by the continuation \eqref{eq:sw42-half-strip} of the thermal two-point function to the middle of the KMS strip. Finally, besides the geometric state $\omega_\beta$, the current admits the primary locally normal KMS states $\omega_q=\omega_\beta\circ\gamma_q$, classified by a real parameter $q$, and in the representations constructed in Section~\ref{sec:scalar} their local modular conjugations happen to coincide with that of the geometric KMS state.

\acknowledgments

We thank David Blanco for his contributions at an early stage of this
project, and Nicol\'as Abate, Mateo Koifman and Yoh Tanimoto for useful
discussions and correspondence. This work was partially supported by
CONICET grant PIP 11220210100685CO.

The ideas and the main computations of this work are due to the authors.
In the final stage of preparation we used the language models GPT-6 Astra
(OpenAI) and Claude Fable 5.2 (Anthropic) to assist with editing, with
consistency checks of the draft, and with the comparison to the recent
literature, in particular the relation to the canonical purification
discussed in Section~\ref{sec:sorce-comparison}, which was verified by
the authors. The authors take full responsibility for the content of
this paper.

\appendix

\section{Generalities: operator algebras, states and modular theory}\label{app:generalities}

This appendix reviews, in a self-contained way, the background material summarized in Section~\ref{sec:setup}: operator algebras and their role in QFT, states and their purifications, thermal (KMS) states, and the modular flow and modular conjugation of Tomita-Takesaki theory.

\subsection{Operator algebras}\label{app:algebras}

In quantum mechanics, subsystems are especially interesting due to the phenomenon of entanglement. A natural generalization of the notion of subsystem is simply a set of operators, for example a set of observables, or a set of unitaries.
If these operators are bounded (which is the same as continuous), then they have no singularities and hence we can take linear combinations, products and adjoints of them without worrying about their domains. In this way we form an algebra, which is in fact a *-algebra because it is self-adjoint, i.e., closed under the operation of taking adjoints. Adding the multiples of the identity operator, the algebra becomes unital (which means containing the identity). We can then take the closure of this algebra in some operator topology. Thus, the consideration of subsystems in quantum mechanics leads us to the study of unital *-algebras of bounded operators which are closed in some operator topology. There are three main operator topologies (associated with the three usual notions of convergence of a sequence of functions), but they only give rise to two classes of algebras: C* and von Neumann algebras.

Let us see this in more detail. Consider a Hilbert space ${\mathcal H}$, and let ${\mathcal B}({\mathcal H})$ be the algebra of all bounded operators on ${\mathcal H}$. A sequence $\{X_n\}\subset {\mathcal B}({\mathcal H})$ is said to converge to an operator $X\in{\mathcal B}({\mathcal H})$ 
\begin{enumerate}[(i)]
    \item uniformly if $\Vert X_n-X\Vert\to 0$, where $\Vert\cdot\Vert$ is the operator norm;
    \item strongly (or pointwise) if $(X_n-X)|\psi\rangle\to 0$ for all $|\psi\rangle\in{\mathcal H}$;
    \item weakly if $\langle\phi|(X_n-X)|\psi\rangle\to 0$ for all $|\phi\rangle,|\psi\rangle\in{\mathcal H}$.
\end{enumerate}
These three notions of convergence are increasingly weak: uniform implies strong, and strong implies weak. A unital *-algebra is called a {\emph{C*-algebra}} if it is closed in the uniform topology, and a {\emph{von Neumann algebra}} if it is closed in the weak topology. Note that the second condition is stronger than the first, so every von Neumann algebra is also a C*-algebra. One might expect that the strong topology would give rise to a third class of algebras, in the middle of these two, but this turns out not to be the case: for a convex set of operators (and in particular for an algebra), the closures in the strong and weak topologies coincide.

An example of a von Neumann algebra is the commutant of any self-adjoint set. The commutant of a set ${\mathcal S}\subseteq{\mathcal B}({\mathcal H})$ is the set ${\mathcal S}'$ of all bounded operators which commute with all operators in ${\mathcal S}$. This is clearly a unital algebra
(it contains the identity and it is closed under linear combinations and products), and one can easily show that it is closed in the weak topology. Moreover, ${\mathcal S}'$ is self-adjoint provided that ${\mathcal S}$ is self-adjoint, so the commutant of a self-adjoint set is indeed a von Neumann algebra. In fact, all von Neumann algebras arise in this way: as famously shown by von Neumann in the so-called bicommutant theorem, every von Neumann algebra ${\mathcal A}$ is equal to its bicommutant, ${\mathcal A}={\mathcal A}''$, so it is the commutant of a self-adjoint set. Two simple properties of commutants are the following: every set is contained in its bicommutant, ${\mathcal S}\subseteq{\mathcal S}''$, and the condition ${\mathcal S}\subseteq{\mathcal T}$ implies ${\mathcal T}'\subseteq{\mathcal S}'$. Applying the second property twice and using the bicommutant theorem we find that any von Neumann algebra containing ${\mathcal S}$ also contains ${\mathcal S}''$. Therefore, the bicommutant of a self-adjoint set is the smallest von Neumann algebra containing that set; it is called the von Neumann algebra generated by the set.

A von Neumann algebra ${\mathcal A}$ is called a factor if its center ${\mathcal A}\cap{\mathcal A}'$ is trivial, i.e., consists only of multiples of the identity. If the Hilbert space factorizes as ${\mathcal H}={\mathcal H}_1\otimes{\mathcal H}_2$, the algebra ${\mathcal B}({\mathcal H}_1)\otimes{\mathds 1}_2$ is a factor, hence the terminology. Factors arising in this way are said to be of type I. In finite dimension all factors are of type I, but in infinite dimension there are other factors. These are classified according to the possibility of defining a non-trivial trace on them, i.e., a trace taking values other than $0$ and $\infty$. A trace on a von Neumann algebra ${\mathcal A}$ is a map $\tau:{\mathcal A}_+\to[0,\infty]$, where ${\mathcal A}_+$ denotes the set of all positive operators in ${\mathcal A}$, which is linear in the appropriate sense{\footnote{${\mathcal A}_+$ is not a vector space but a convex cone, i.e., a set closed under linear combinations with positive coefficients. The linearity of $\tau$ means that it preserves this type of linear combinations.}}, assigns non-zero values to non-zero operators and satisfies $\tau(a^\dagger a)=\tau(aa^\dagger)$. Of course, the standard trace satisfies these properties. On a factor of type I, the standard trace may act trivially (i.e., it may only take the values $0$ and $\infty$; this is what happens when ${\mathcal H}_2$ is infinite-dimensional), but it is always possible to define a non-trivial trace: take for example $\tau(X_1\otimes{\mathds 1}_2)=\Tr X_1$. Factors which are not of type I but still admit a non-trivial trace are said to be of type II, and the remaining factors are said to be of type III (for a nice account on this see \cite{Sorce:2023fdx}).

A vector $|\psi\rangle\in{\mathcal H}$ is said to be cyclic for a set of operators ${\mathcal S}\subseteq{\mathcal B}({\mathcal H})$ if the set ${\mathcal S}|\psi\rangle$ is dense in ${\mathcal H}$, i.e., if the closure of this set is the whole Hilbert space; $|\psi\rangle$ is said to be separating for ${\mathcal S}$ if $s|\psi\rangle\ne 0$ for every non-zero $s\in{\mathcal S}$. Cyclic for a set implies separating for the commutant: for $s'\in{\mathcal S}'$, the condition $s'|\psi\rangle=0$ implies that $s'$ annihilates all vectors in the set ${\mathcal S}|\psi\rangle$, and hence vanishes if that set is dense. In the case of a unital *-algebra ${\mathcal A}$, the converse statement is also true: separating for ${\mathcal A}'$ implies cyclic for ${\mathcal A}$. Indeed, the orthogonal projection $P$ onto the orthogonal complement of ${\mathcal A}|\psi\rangle$ satisfies{\footnote{The first property uses that ${\mathcal A}$ is a *-algebra; to see it, note that ${\mathcal A}$ leaves the subspace $({\mathcal A}|\psi\rangle)^\perp$ invariant, and that the projection onto any invariant closed subspace commutes with all operators in the algebra. The second property uses that ${\mathcal A}$ is unital.}} $P\in{\mathcal A}'$ and $P|\psi\rangle=0$, so if $|\psi\rangle$ is separating for ${\mathcal A}'$ we have $P=0$ and hence ${\mathcal A}|\psi\rangle$ is dense. Thus, a vector is cyclic for a unital *-algebra if and only if it is separating for its commutant.

In QFT, every open spacetime region has associated a von Neumann algebra. Consider for example the case of a scalar field theory. The field $\phi$ is not an operator-valued function but an operator-valued distribution, so it assigns operators to test functions rather than to spacetime points. Given a test function $f$, the operator $\phi(f)$ is generally unbounded, but, assuming that $\phi$ is real and restricting to real test functions, $e^{i\phi(f)}$ is bounded\footnote{for a more detailed account with different but equivalent ways of quantization of the free field $\phi$ see for example 
\cite{Garbarz:2021rqu}}. The von Neumann algebra associated with an open region ${\mathcal O}$ is\footnote{We will loosely say that $\phi(x)$ or $\phi(f)$ belongs to $\mathcal{A}_{\mathcal O}$ if $x\in \mathcal{O}$ or supp$(f)\subset\mathcal
O$, respectively. However the correct statement is that  $e^{i\phi(f)}$ is in the von Neumann algebra $\mathcal{A}_{\mathcal O}$ and $\phi(f)$ is \textit{affiliated} with such algebra, if supp$(f)\subset\mathcal
O$. A closed operator $s$ on a Hilbert space is affiliated with $\mathcal{A}$ if $\mathcal{A}'D(s)\subseteq{D(s)}$ and $a's\subseteq sa' $ for all $a' \in \mathcal{A}'$. So, this essentially says that even though $\phi(f)$ does not belong to $\mathcal{A}_{\mathcal O}$, it  commutes with all elements of $\mathcal{A}_{\mathcal O}'$.}
\begin{equation}
    {\mathcal A}_{\mathcal O}=\{e^{i\phi(f)},\,\supp(f)\subset{\mathcal O}\}'',
\end{equation}
where $\supp(f)$ denotes the support of $f$. Note that ${\mathcal A}_{{\mathcal O}_1}\subseteq{\mathcal A}_{{\mathcal O}_2}$ if ${\mathcal O}_1\subseteq{\mathcal O}_2$, and that algebras associated with spacelike separated regions commute\footnote{We are implicitly assuming that $\phi(f)$ behaves as a Wightman field, so we have $[\phi(f),\phi(g)]=0$ if $f$ and $g$ are supported on spacelike separated regions.}, ${\mathcal A}_{{\mathcal O}_1}\subseteq{\mathcal A}_{{\mathcal O}_2}'$ if ${\mathcal O}_1\subseteq{\mathcal O}_2'$, where ${\mathcal O}'$ denotes the causal complement of ${\mathcal O}$, i.e., the union of all open regions spacelike separated from ${\mathcal O}$. In the algebraic approach to QFT, this assignment of algebras to regions is viewed as the essential feature of the theory. That is, a QFT is viewed as an assignment of algebras to regions satisfying the above properties, among others. To be more precise, we will be interested in a net of algebras called conformal net on $\mathbb{R}^{1,1}$, which is  the algebraic setup of a conformal field theory in two-dimensional Minkowski spacetime, and will be introduced in Section \ref{2dCFTs}.

As it turns out, the local algebras of QFT are generically type III factors, with the exception of the algebras associated with the empty region and the entire spacetime, which are type I (they are ${\mathbb C}{\mathds 1}$ and ${\mathcal B}({\mathcal H})$ respectively). 
Another important result in this context is the Reeh-Schlieder theorem, which says that the vacuum vector is cyclic for the algebra associated with any non-empty region, no matter how small. For a region ${\mathcal O}$ such that ${\mathcal O}'$ is non-empty, the vacuum is cyclic for ${\mathcal A}_{{\mathcal O}'}$, hence separating for ${\mathcal A}_{{\mathcal O}'}'$ and in particular for ${\mathcal A}_{\mathcal O}\subseteq{\mathcal A}_{{\mathcal O}'}'$. Thus, the Reeh-Schlieder theorem implies that the vacuum is separating for any region with a non-empty causal complement.

\subsection{States}
\label{sec:states}

A state on a unital *-algebra ${\mathcal A}$ is a linear map $\omega:{\mathcal A}\to{\mathbb C}$ which is positive, $\omega(a^\dagger a)\ge 0$, and normalized, $\omega({\mathds 1})=1$. These properties alone imply{\footnote{This is especially easy to see in the case where $a$ is Hermitian, because in this case we have $-\Vert a\Vert\le a\le \Vert a\Vert$.}} 
$|\omega(a)|\le\Vert a\Vert$, so every state is continuous in the uniform topology. What we usually call a state in quantum mechanics, namely a density matrix $\rho$, defines a state in this algebraic sense via the equation $\omega(a)=\Tr(\rho a)$. In finite dimension all states arise in this way, but in infinite dimension there are other states. In the case where ${\mathcal A}$ is a von Neumann algebra, the states arising from a density matrix are called normal, and  they are precisely the states which are continuous in the weak topology\footnote{Sometimes in the literature they are defined as continuous with respect to the $\sigma$-weak topology, which is finer than the weak-topology, but on the unit ball it coincides with the weak topology \cite{kadison1997fundamentalsII}. There are also other equivalent definitions, see  \cite{BR1}.} on the unit ball $\{a\in{\mathcal A},\ \Vert a\Vert\le 1\}$. In particular, every normal state $\omega$ is sequentially continuous in the weak topology, i.e., $\omega(a_n)\to \omega(a)$ if $a_n\to a$ weakly, because all weakly convergent sequences are bounded in norm{\footnote{In first-countable topologies, sequential continuity is equivalent to continuity, but the weak topology is not first-countable (neither is the strong topology). The fact that normal states may not be weakly continuous outside the unit ball has consequences when considering nets instead of sequences (a net is a generalized sequence, where the index is not necessarily a natural number but an element of a more general set which retains some of the ordering properties of the natural numbers). The image by a normal state of a weakly convergent net may not be convergent, because weakly convergent nets are not necessarily bounded in norm.\label{sequential}}}.

Two more comments about normal states are in order. First, note that the density matrix is not required to be in the algebra. For example, a factor of type III contains no density matrices (because there is no way to define a trace on it taking values other than $0$ and $\infty$), but still admits plenty of normal states. And second, two different density matrices may give rise to the same state. For example, if ${\mathcal H}={\mathcal H}_1\otimes{\mathcal H}_2$ the density matrices $\rho_1\otimes\rho_2$ and $\rho_1\otimes\sigma_2$ define the same state on ${\mathcal B}({\mathcal H}_1)\otimes{\mathds 1}_2$. This does not happen if the algebra is ${\mathcal B}({\mathcal H})$: in this case, every normal state has associated a unique density matrix.


Every state can be purified. That is, for every state $\omega$ on a unital *-algebra ${\mathcal A}$ there is an auxiliary Hilbert space ${\mathcal K}$, a representation $\pi$ of ${\mathcal A}$ by bounded operators on ${\mathcal K}$, 
$\pi:{\mathcal A}\to{\mathcal B}({\mathcal K})$ 
and a vector $|\psi\rangle\in{\mathcal K}$ such that $\omega(a)=\langle\psi|\pi(a)|\psi\rangle$. Moreover, these objects can be chosen so that $|\psi\rangle$ is cyclic for $\pi({\mathcal A})$.
In the case where $\omega$ arises from a density matrix, this result is intuitive: we know that any density matrix $\rho\in{\mathcal B}({\mathcal H})$ can be purified by a vector in ${\mathcal H}\otimes{\mathcal H}$, and if this vector is not cyclic we can make it cyclic by discarding vectors from the auxiliary Hilbert space.
The general result, valid for arbitrary states, is proved by an explicit construction of the auxiliary Hilbert space, the representation and the cyclic vector called the GNS construction (after Gel'fand, Naimark and Segal) \cite{haag1996local}.

Let us review the GNS construction in the simple case where $\omega$ is faithful, i.e., $\omega(a^\dagger a)\ne 0$ for $a\ne 0$. In this case the equation $\langle a,b\rangle=\omega(a^\dagger b)$ defines an inner product on ${\mathcal A}$. Take ${\mathcal K}$ to be the completion of ${\mathcal A}$ in the topology associated with this inner product. By construction, ${\mathcal A}$ is a dense subspace of ${\mathcal K}$. Take now $\pi(a)$ to be the operator defined in this dense subspace by $\pi(a)b=ab$. The property $|\omega(a)|\le\Vert a\Vert$ ensures that this operator is bounded, so it can be extended to the whole Hilbert space by continuity. It is straightforward to see that $\pi$ is a representation, i.e., it preserves all the operations of the algebra and maps the identity to itself. Finally, take $|\psi\rangle={\mathds 1}$. Clearly, this vector is cyclic, because $\pi({\mathcal A})|\psi\rangle={\mathcal A}$, and we have $\langle\psi|\pi(a)|\psi\rangle=\langle {\mathds 1},a\rangle=\omega(a)$, as desired.

So far we have only discussed the existence of a cyclic purification of 
any given state. In fact, this purification is unique up to unitary equivalence. That is, if $({\mathcal K}_1,\pi_1,|\psi_1\rangle)$ and $({\mathcal K}_2,\pi_2,|\psi_2\rangle)$ are two cyclic purifications of a state $\omega$, then there is a unitary operator $U:{\mathcal K}_1\to{\mathcal K}_2$ such that $\pi_2(a)=U\pi_1(a)U^\dagger$ and $|\psi_2\rangle=U|\psi_1\rangle$. To see this, consider the operator $U$ defined by
\begin{equation}
    U\pi_1(a)|\psi_1\rangle=\pi_2(a)|\psi_2\rangle.
\end{equation}
This operator is well-defined (in the sense of being single-valued) because the vectors $\pi_1(a)|\psi_1\rangle$ and $\pi_2(a)|\psi_2\rangle$ have the same norm, and hence if one of them vanishes the other vanishes too. For the same reason, $U$ is an isometry. The cyclicity of $|\psi_1\rangle$ and $|\psi_2\rangle$ implies that this isometry has dense domain and range, so it can be extended by continuity to the whole Hilbert space giving a unitary operator. This operator obviously satisfies $U|\psi_1\rangle=|\psi_2\rangle$, and moreover
\begin{alignat}{2}
U\pi_1(a)U^\dagger\pi_2(b)|\psi_2\rangle&=U\pi_1(a)\pi_1(b)|\psi_1\rangle=U\pi_1(ab)|\psi_1\rangle=\pi_2(ab)|\psi_2\rangle\nonumber\\
&=\pi_2(a)\pi_2(b)|\psi_2\rangle,
\end{alignat}
so $U\pi_1(a)U^\dagger=\pi_2(a)$, as we wanted to show.

The purifications (cyclic or not) of faithful states are characterized by involving a faithful representation (i.e., an injective one) and a separating vector. Indeed, if $\omega$ is faithful and $({\mathcal K},\pi,|\psi\rangle)$ is a purification, then $\pi(a)|\psi\rangle\ne 0$ for $a\ne 0$, which implies that $\pi$ is faithful and $|\psi\rangle$ is separating for $\pi({\mathcal A})$. Conversely, if $\omega$ admits a purification $({\mathcal K},\pi,|\psi\rangle)$ with $\pi$ faithful and $|\psi\rangle$ separating then it is faithful, because for $a\ne 0$ we have $\pi(a)|\psi\rangle\ne 0$ and hence $\omega(a^\dagger a)\ne 0$.


In the case where ${\mathcal A}$ is a von Neumann algebra and $\omega$ is a normal state, the image $\pi({\mathcal A})$ of the GNS representation 
(or any other representation in which $\omega$ can be purified by a cyclic vector) is also a von Neumann algebra \cite{BR1}. If, besides being normal, $\omega$ is faithful, then $\pi$ is an isomorphism between von Neumann algebras, which, as it turns out, implies that it restricts to a homeomorphism (i.e., a continuous map with continuous inverse) between the unit balls, both in the strong and the weak topologies. On the other hand, if a given representation of ${\mathcal A}$ contains a purifying vector for a faithful normal state (which requires the representation to be faithful), 
then it contains purifying vectors for all normal states.
This is easy to see when the algebra is ${\mathcal B}({\mathcal H})$ and ${\mathcal H}$ is finite-dimensional. Indeed, a normal state on ${\mathcal B}({\mathcal H})$ is faithful if and only if the associated density matrix is injective, which in finite dimension implies invertible{\footnote{In infinite dimension, an injective operator needs not be surjective, and in fact a density matrix can never be surjective. This is because its eigenvalues have to tend quickly to zero to make the trace finite, and therefore its image consists of vectors whose components in an eigenbasis tend quickly to zero.}}. If an invertible density matrix $\rho$ is purified in a representation $\pi$ by a vector $|\psi\rangle$, then an arbitrary density matrix $\sigma$ is purified in the same representation by $\pi(\sigma^{1/2}\rho^{-1/2})|\psi\rangle$.

The above result has an immediate consequence in QFT: if a region ${\mathcal O}$ has a non-empty causal complement, then all normal states on ${\mathcal A}_{\mathcal O}$
can be purified in the identity representation, i.e., for every normal state $\omega$ on 
this algebra 
there is a vector $|\psi\rangle\in{\mathcal H}$ such that $\omega(a)=\langle\psi|a|\psi\rangle$. This is because the state  $a\mapsto\langle\Omega|a|\Omega\rangle$ induced by the vacuum $|\Omega\rangle$, which obviously is normal and can be purified in the identity representation, is faithful by the Reeh-Schlieder theorem.

\subsection{Thermal states}\label{app:thermal}

In quantum statistical mechanics, a thermal state is usually defined as a density matrix of the form
\begin{equation}\label{thermalrho}
    \rho=\frac{e^{-\beta H}}{Z},
\end{equation}
where $H$ is the Hamiltonian of the system, $\beta$ is a positive number (the inverse temperature) and $Z$ is a normalization factor, the partition function. Typically, the partition function grows exponentially with the volume of the system, because its logarithm, the free energy, is an extensive quantity. Therefore, in the infinite-volume limit (the limit in which one expects to reproduce thermodynamics), the partition function diverges, which means that $e^{-\beta H}$ fails to be trace-class and hence that the thermal density matrix is ill-defined.

Of course, this does not prevent us from making predictions about the infinite-volume limit. The usual way to proceed is to start at finite volume, compute expectation values and then take the infinite-volume limit of those expectation values. This limit does not exist for all observables: for example, the energy and other extensive quantities diverge{\footnote{The Hamiltonian is usually not a bounded operator, but even bounded functions of $H$ do not generally have well-defined thermal expectation values in the infinite-volume limit.}}. But, for observables localized in bounded regions, the limit usually exists. This gives rise to an algebraic state $\omega$ on the algebra of all operators localized in bounded regions, which cannot be expressed in terms of a density matrix. Since any algebraic state is continuous in the uniform topology, we can extend $\omega$ to the uniform closure of this algebra, which is a C*-algebra called the quasilocal algebra. 

The state $\omega$ can be characterized directly in the infinite-volume limit, without having to resort to the limiting procedure described above. This is done via the so-called KMS condition (after Kubo, Martin and Schwinger). Let ${\mathcal A}$ be a C*-algebra, and let $\alpha:{\mathbb R}\times{\mathcal A}\to{\mathcal A}$ be a one-parameter group of automorphisms of ${\mathcal A}$ (i.e., isomorphisms from the algebra to itself), that to each $t\in{\mathbb R}$ assigns the automorphism $\alpha_t$. A state $\omega$ on ${\mathcal A}$ is said to satisfy the KMS condition at inverse temperature $\beta$ with respect to $\alpha$
if, for every $a,b\in{\mathcal A}$, there is a bounded function $F_{ab}$ on the strip $\{z\in{\mathbb C}\,|\,\Im z\in[0,\beta]\}$,
analytic on the interior of the strip and continuous on its boundary, such that, for $t\in{\mathbb R}$,
\begin{equation}\label{kms}
    F_{ab}(t)=\omega(a\alpha_t(b))\qquad F_{ab}(t+i\beta)=\omega(\alpha_t(b)a).
\end{equation}
The same definition applies for $\beta<0$, with the strip $\{z\in{\mathbb C}\,|\,\Im z\in[\beta,0]\}$; this will be used for the modular group in Appendix~\ref{app:modular}.
As we will now see, when the thermal density matrix is well-defined, the corresponding algebraic state on ${\mathcal B}({\mathcal H})$ satisfies the KMS condition with respect to the time-translation group $\alpha_t(X)=e^{itH}Xe^{-itH}$, and there are no other density matrices with this property. Thus, the KMS condition completely characterizes thermal states. Moreover, the condition makes no reference to density matrices, so it can be taken as the definition of a thermal state when the thermal density matrix is not well-defined. Under reasonable assumptions, this definition agrees with the limiting procedure described above \cite{Haag:1967sg}.

Let us see that, if the thermal density matrix is well-defined, the corresponding state on ${\mathcal B}({\mathcal H})$ satisfies the KMS condition with respect to time translations. Consider first an operator $A$ of a somewhat special type, namely a finite linear combination
of operators of the form $|m\rangle\langle n|$, with $|m\rangle$ and $|n\rangle$ eigenvectors of the Hamiltonian. The function $t\mapsto \alpha_t(A)=e^{itH}Ae^{-itH}$ extends to an analytic function ${\mathbb C}\to{\mathcal B}({\mathcal H})$, and hence the function
\begin{equation}
    F_{XA}(z)=\frac{1}{Z}\Tr(e^{-\beta H}Xe^{izH}Ae^{-izH}),
\end{equation}
with $X$ an arbitrary bounded operator, is analytic on the whole complex plane. Moreover, it clearly satisfies the boundary condition (\ref{kms}), and it is bounded in any region where $\im z$ is bounded, because
\begin{alignat}{2}
    |F_{XA}(z)|&\le\Vert Xe^{izH}Ae^{-izH}\Vert\le\Vert X\Vert\Vert e^{izH}Ae^{-izH}\Vert\nonumber\\
    &=\Vert X\Vert\Vert e^{-(\im z) H}Ae^{(\im z)H}\Vert.
\end{alignat}
In particular, it is bounded on the strip $\Im z\in[0,\beta]$, and hence satisfies all the properties required in the KMS condition. In fact, by Hadamard's three lines theorem, the analyticity and boundedness, together with the boundary condition (\ref{kms}), imply
\begin{equation}\label{uniformbound}
    |F_{XA}(z)|\le\sqrt{\omega(\{X,X^\dagger\})\,\omega(\{A,A^\dagger\})}
\end{equation}
for $z$ in the strip, where $\{\,,\}$ denotes the anticommutator. Now, for any bounded operator $Y$, the operator $Y_N=P_NYP_N$, where $P_N$ denotes the projection onto the subspace spanned by the first $N$ eigenvectors of $H$, is of the type considered above (a finite linear combination of operators of the form $|m\rangle\langle n|$), and the sequences $\{Y_N\}$ and $\{Y_{N}^\dagger\}$ converge strongly to $Y$ and $Y^\dagger$ respectively.
The bound (\ref{uniformbound}), together with the continuity properties of normal states, implies that the sequence $\{F_{XY_{N}}\}$ is uniformly Cauchy, and hence converges uniformly. Therefore, it converges to a function which is
bounded on the strip, analytic on its interior and continuous on its boundary. Moreover, this function satisfies the boundary condition (\ref{kms}) for the operators $X$ and $Y$, by the continuity properties of normal states, so the KMS condition is indeed satisfied for arbitrary pairs of operators.

Let us now see that the converse statement also holds, namely that, if a normal state on ${\mathcal B}({\mathcal H})$ satisfies the KMS condition with respect to time translations, then the corresponding density matrix is the thermal one. To see this, note first from the KMS condition that, if $b$ is invariant under $\alpha$, then $\omega(ab)=\omega(ba)$. If the algebra is ${\mathcal B}({\mathcal H})$ and $\omega$ is normal, this implies that the associated density matrix $\rho$ commutes with all $\alpha$-invariant operators. Taking $\alpha$ to be the group of time translations, we obtain in particular that $\rho$ commutes with $H$, and hence there is a simultaneous eigenbasis of both operators (assuming such basis exists, see \cite{haag1996local} for a more general treatment). 
Applying the KMS condition to the operators $|m\rangle\langle n|$ and $|n\rangle\langle m|$, where $|m\rangle$ and $|n\rangle$ are arbitrary elements of this eigenbasis, one finally obtains that $\rho$ has the form (\ref{thermalrho}).

An important property of KMS states is that they are invariant under $\alpha$. Indeed, the boundary condition (\ref{kms}) implies $F_{{\mathds 1}a}(t)=F_{{\mathds 1}a}(t+i\beta)$. Extending $F_{{\mathds 1}a}$ to the entire complex plane by periodicity, one obtains a function which is bounded and analytic everywhere (because  
a function analytic on both sides of a segment and continuous at the segment is analytic also at the segment), 
and hence, by Liouville's theorem, it is constant. In particular it is constant on the real line, so $\omega(\alpha_{t}(a))=\omega(a)$, as we wanted to show.

Given a C*-algebra and an automorphism group, it is not guaranteed that a KMS state will exist.
And if it does exist at a certain temperature, it is not necessarily unique. The non-uniqueness means that we have different equilibrium states at the same temperature, which signals the presence of a phase transition. The set of KMS states at a given temperature is clearly convex (if $\omega_1$ and $\omega_2$ are KMS at inverse temperature $\beta$, then so is any convex combination of $\omega_1$ and $\omega_2$). The phases are identified with the extreme points of this convex set, the remaining points corresponding to mixtures of phases.

\subsection{Modular flow and modular conjugation}\label{app:modular}

Every faithful normal state on ${\mathcal B}({\mathcal H})$ satisfies the KMS condition at unit temperature with respect to some automorphism group. Indeed, the density matrix $\rho$ associated with a faithful normal state is injective and hence can be written in the form $\rho=e^{-H}/Z$ for some Hamiltonian $H$. The KMS condition at unit temperature is then satisfied with respect to the automorphism group generated by this Hamiltonian, $X\mapsto\rho^{-it}X\rho^{it}$. 

In fact, there are no other automorphism groups with this property. Indeed, all automorphisms of ${\mathcal B}({\mathcal H})$ are inner, i.e., for every automorphism $\alpha$
there is a unitary operator $U$ such that 
$\alpha(X)=UXU^\dagger$. Note that this operator is not unique because it can be multiplied by an arbitrary phase. If $\alpha:t\mapsto\alpha_t$ is a one-parameter group of automorphisms, then for each $t$ we can choose the phase so that the associated map $U:t\mapsto U_t$ is also a one-parameter group. Now, if the state satisfies the KMS condition with respect to $\alpha$, then in particular it is invariant under $\alpha$, so $U_t$ commutes with the associated density matrix $\rho$. In fact, applying the KMS condition to a pair of operators such that the second commutes with $\rho$,
one sees that $U_t$ commutes with all operators commuting with $\rho$, i.e., $U_t\in\{\rho\}''$. The bicommutant of a set consisting of a single operator is the algebra of all bounded functions of that operator, so we conclude that $U_t$ is a function of $\rho$. This means that every eigenvector of $\rho$ is also an eigenvector of $U_t$; the group property of $U$ then implies that the corresponding eigenvalue has the form $e^{iEt}$, with $E$ a constant. Finally, if the KMS condition is satisfied at unit temperature, applying it to the operators $|m\rangle\langle n|$ and $|n\rangle\langle m|$, with $|m\rangle$ and $|n\rangle$ arbitrary elements of an eigenbasis of $\rho$, we find that $U_t=\rho^{-it}$, as we wanted to show.

A highly non-trivial fact is that the above result extends to arbitrary von Neumann algebras: for every faithful normal state $\omega$ on a von Neumann algebra ${\mathcal A}$, there is a unique one-parameter group of automorphisms of ${\mathcal A}$ with respect to which $\omega$ satisfies the KMS condition at unit temperature \cite{BRII}. Following the standard convention, we call modular group, or modular flow, of $\omega$ this group run backwards, $t\mapsto\alpha_t$. With respect to it $\omega$ satisfies the KMS condition at $\beta=-1$: $F_{ab}$ is analytic on the strip $-1<\Im z<0$, with $F_{ab}(t)=\omega(a\alpha_t(b))$ and $F_{ab}(t-i)=\omega(\alpha_t(b)a)$. For ${\mathcal B}({\mathcal H})$ it is $\alpha_t(X)=\rho^{it}X\rho^{-it}$. Note that what carries over from ${\mathcal B}({\mathcal H})$ to arbitrary smaller algebras is the existence and uniqueness of an automorphism group with certain properties,
not the explicit form of that group. For type I factors the form is similar{\footnote{This is not surprising, because a type I factor is essentially the same as ${\mathcal B}({\mathcal H})$.}}, 
$\alpha_t(X_1\otimes{\mathds 1}_2)=\rho_{1}^{it}X_1\rho_{1}^{-it}\otimes{\mathds 1}_2$, where $\rho_1$ denotes the reduced density matrix, obtained by taking a partial trace of the global density matrix. But for other algebras the modular group is more complicated. 

Let $\omega$ be a faithful normal state on a von Neumann algebra ${\mathcal A}$, and let $\alpha$ be the corresponding modular group. It is not hard to show that the map $t\mapsto\alpha_t(a)$ is continuous in the strong topology. We will say that $a$ is analytic with respect to $\alpha$ if this map extends to an analytic function ${\mathbb C}\to{\mathcal A}$, which we will denote as $z\mapsto\alpha_z(a)$. Note that the identity is analytic, and that linear combinations and products of analytic operators are also analytic. Moreover, if $a$ is analytic, then $f(z)=[\alpha_{z^*}(a)]^\dagger$ is an analytic function satisfying $f(t)=\alpha_t(a^\dagger)$ for $t\in{\mathbb R}$, so $a^\dagger$ is also analytic (with $\alpha_z(a^\dagger)=[\alpha_{z^*}a]^\dagger$). Therefore, the analytic operators form a unital *-subalgebra ${\mathcal A}_\alpha\subseteq{\mathcal A}$. This subalgebra is dense in ${\mathcal A}$ in the strong topology, and in fact sequentially dense{\footnote{A similar comment to footnote \ref{sequential} applies here: in first-countable topologies, sequentially dense is equivalent to just dense, but in general topologies it is a stronger condition.}}: every $a\in{\mathcal A}$ is the strong limit of a sequence $\{a_n\}\subset{\mathcal A}_\alpha$. Indeed, take
\begin{equation}
    a_n=\sqrt{\frac{n}{\pi}}\int_{-\infty}^\infty dt\,\alpha_t(a)e^{-nt^2}.
\end{equation}
One can easily show that $a_n$ is analytic and that the sequence converges strongly to $a$ \cite{BR1}. For $a$ and $b$ analytic, the KMS condition simply says that
\begin{equation}\label{kmsanalytic}
    \omega(\alpha_{i/2}(a)\alpha_{-i/2}(b))=\omega(ba),
\end{equation}
where we have used the invariance of $\omega$ under $\alpha$.

Consider now a cyclic purification $({\mathcal K},\pi,|\psi\rangle)$ of $\omega$. By the continuity properties of $\pi$,
the algebra $\pi({\mathcal A}_\alpha)$ is dense in $\pi({\mathcal A})$, which implies that the subspace $\pi({\mathcal A}_\alpha)|\psi\rangle$ is dense in ${\mathcal K}$. In other words, $|\psi\rangle$ is cyclic also for this smaller algebra; note that it is also separating due to the faithfulness of $\omega$. The latter property allows us to define an operator $J$ on $\pi({\mathcal A}_\alpha)|\psi\rangle$ by the equation 
\begin{equation}\label{J}
    J\pi(a)|\psi\rangle=\pi(\alpha_{-i/2}(a^\dagger))|\psi\rangle.
\end{equation}
This is an antilinear operator which, by the KMS condition (\ref{kmsanalytic}), satisfies $J^\dagger J={\mathds 1}$ (recall that the adjoint of an antilinear operator $A$ is defined with a complex conjugation with respect to the case of linear operators, $\langle\phi| A^\dagger|\chi\rangle=\langle\chi|A|\phi\rangle$). Since both its domain and range are dense, $J$ extends by continuity to an antiunitary operator, which is called the {\emph{modular conjugation}} associated with the purification $({\mathcal K},\pi,|\psi\rangle)$. 
Note from the definition (\ref{J}) that $J|\psi\rangle=|\psi\rangle$ and $J^2={\mathds 1}$ (which, together with antiunitarity, implies that $J$ is self-adjoint and antilinear). Another easy property is that, for $a$ and $b$ analytic,
\begin{equation}
    J\pi(a)J\pi(b)|\psi\rangle=\pi(b)J\pi(a)|\psi\rangle,
\end{equation}
which implies that $J\pi(a)J$ commutes with $\pi(b)$, 
\begin{equation}
J\pi(a)J\pi(b)\pi(c)|\psi\rangle=\pi(b)\pi(c)J\pi(a)|\psi\rangle=\pi(b)J\pi(a)J\pi(c)|\psi\rangle.
\end{equation}
In other words, $J\pi({\mathcal A}_\alpha)J\subseteq\pi({\mathcal A}_\alpha)'=\pi({\mathcal A})'$. Taking closures we find $J\pi({\mathcal A})J\subseteq\pi({\mathcal A})'$.
In fact, one can show that the opposite inclusion also holds, which leads to the well-known result
\begin{equation}\label{TTth}
    J\pi({\mathcal A})J=\pi({\mathcal A})'.
\end{equation}
Note finally that the modular conjugations associated with two different cyclic purifications of $\omega$ are unitarily related, $J_2=UJ_1U^\dagger$, where $U:{\mathcal K}_1\to {\mathcal K}_2$ is the unitary operator relating both purifications. We will use this fact frequently. 

The presentation so far did not make contact with the usual presentation of the Tomita-Takesaki modular theory. Given a von Neumann algebra $\mathcal{A}$ and a cyclic and separating vector $|\psi\rangle$, then the Tomita operator $S$ is defined by,
\begin{equation}\label{Tomitaoperator}
    S\pi(a)|\psi\rangle=\pi(a^\dagger)|\psi\rangle.
\end{equation}
with polar decomposition,
\begin{equation}
    S=J \Delta^{1/2}.
\end{equation}
In \eqref{Tomitaoperator} the domain of $S$ is $\pi(\mathcal{A})|\psi\rangle$ and one actually has to take closure of $S$ (we keep calling it $S$). The connection with the modular group $\alpha_t$ is that it is implemented by the adjoint action generated by the (log of the ) positive self-adjoint operator $\Delta$, namely
\begin{equation}
    \pi(\alpha_t (a))=\Delta^{it}\pi(a)\Delta^{-it},
\end{equation} 
with $\Delta|\psi\rangle=|\psi\rangle$. Then it is easy to see where the definition \eqref{J} comes from, on analytic operators:
\begin{equation}
    J\pi(a)|\psi\rangle=JS\pi(a^\dagger)|\psi\rangle=J^2\Delta^{1/2}\pi(a^\dagger)|\psi\rangle.
\end{equation}
which as long as $J^2=\mathds 1$, gives \eqref{J}. Actually, it is possible to prove that a $J$ given by \eqref{J} must be the one in the polar decomposition of $S$, by a characterization\footnote{An operator $J$ is the modular conjugation associated to a von Neumann algebra $R$ and a cyclic and separating vector $|\psi\rangle$ if the following conditions hold: (i) $(J|\phi\rangle,J|\varphi\rangle )=(|\varphi\rangle,|\phi\rangle )$ for all $|\varphi\rangle$ and $\phi\rangle$ in the Hilbert space, (ii) $J^2=\mathds  1$, (iii) $J R J=R' $, (iv) $J|\psi\rangle=|\psi\rangle$, and (v) $(|\psi\rangle,Q J Q J|\psi\rangle)\geq 0$ for all $Q\in R$, with equality iff $Q=0$. The operator $J$ defined by \eqref{J} satisfies all these properties.} of the modular conjugation operator (see \cite{Araki:1972zza}). 

To familiarize with the modular conjugation, consider the following representation of ${\mathcal B}({\mathcal H})$. The representation space is the space of all Hilbert-Schmidt operators on ${\mathcal H}$, namely the operators $s$ such that $s^\dagger s$ is trace-class. This is a Hilbert space with inner product $\langle s,t\rangle=\Tr(s^\dagger t)$, and also a two-sided *-ideal of ${\mathcal B}({\mathcal H})$ (i.e., it is self-adjoint and closed under multiplication by arbitrary bounded operators, both from the left and from the right). The representation itself is left multiplication,
$\pi(X)s=Xs$. Any normal state on ${\mathcal B}({\mathcal H})$ can be purified in this representation, in several ways: if $\rho$ is the corresponding density matrix, any operator $s$ with $ss^\dagger=\rho$ (which is automatically Hilbert-Schmidt) is a purifying vector, 
\begin{equation}
    \langle s,\pi(X)s\rangle=\langle s,Xs\rangle=\Tr(s^\dagger Xs)=\Tr(\rho X).
\end{equation}
The standard choice is $s=\sqrt{\rho}$. If the state is faithful ($\rho$ injective), ${\mathcal B}({\mathcal H})\sqrt\rho$ is dense and hence the standard purification is cyclic. Let us compute the modular conjugation in this case. The modular flow is $\alpha_t(X)=\rho^{it}X\rho^{-it}$, so any operator of the form $|i\rangle\langle j|$, with $|i\rangle$ and $|j\rangle$ eigenvectors of $\rho$, is analytic. Applying the definition (\ref{J}) to these operators we obtain $J|i\rangle\langle j|=|j\rangle\langle i|$ and hence
\begin{equation}\label{JHilbertSchmidt}
    Js=s^\dagger.
\end{equation}
Thus, the modular conjugation is a very simple operator when the algebra is ${\mathcal B}({\mathcal H})$ and the purification is the standard one in the space of Hilbert-Schmidt operators. From the above equation it follows that
\begin{equation}
    J\pi(X)Js=sX^\dagger.
\end{equation}
Another representation of ${\mathcal B}({\mathcal H})$ in which all normal states can be purified, and which is perhaps more familiar than the one described above, is $\pi(X)=X\otimes{\mathds 1}$ on ${\mathcal H}\otimes{\mathcal H}$. Both representations are unitarily equivalent, although not in a natural way, i.e., there is no preferred unitary operator relating them. Given an orthonormal basis $\{|i\rangle\}$ of ${\mathcal H}$, the unitary operator defined by $|i\rangle\langle j|\mapsto |i\,j\rangle$ relates these two representations. Under this unitary, the modular conjugation (\ref{JHilbertSchmidt}) is mapped to $J|i\,j\rangle=|j\,i\rangle$, which satisfies $J(|i\rangle\langle j|\otimes{\mathds 1})J={\mathds 1}\otimes|i\rangle\langle j|$.

Going back to the general case, since the representation $\pi$ contains a purifying vector for a faithful normal state (the vector $|\psi\rangle$, which purifies the state $\omega$), we know that it contains purifying vectors for all normal states. In fact, there is a subset of the representation space ${\mathcal K}$ which already has this property: the closure ${\mathcal P}$ of the set of all vectors of the form $\pi(a)J\pi(a)|\psi\rangle$, with $a\in{\mathcal A}$. This set turns out to be a convex cone, and it is called the natural positive cone associated with $\omega$. Some of its properties are the following \cite{BR1} (a larger list is given in Appendix~\ref{app:cone}):
\begin{enumerate}[(i)]
    \item every normal state $\varphi$ on ${\mathcal A}$ is purified in the representation $\pi$ by a unique vector $|\Psi_\varphi\rangle\in{\mathcal P}$;
    \item if $\varphi$ is also faithful, then $|\Psi_\varphi\rangle$ is cyclic for $\pi({\mathcal A})$;
    \item the corresponding modular conjugation is independent of $\Psi_\varphi$.
\end{enumerate}
This is easy to see when the algebra is ${\mathcal B}({\mathcal H})$ and the purification is the standard one described above. In that case, one can see from (\ref{JHilbertSchmidt}) that ${\mathcal P}$ is the set of all positive Hilbert-Schmidt operators. Among all the possible purifications of a density matrix $\sigma$, only one is in ${\mathcal P}$, namely the standard one, $\sqrt\sigma$. As explained above, this purification is cyclic provided that $\sigma$ is injective, and the associated modular conjugation is given by (\ref{JHilbertSchmidt}) independently of $\sigma$.

\section{The positive cone}\label{app:cone}

We follow \cite{BR1} and \cite{Summers:2005xj} with slight variations. The natural positive cone associated to a von Neumann algebra $\mathcal{A}$ and a cyclic and separating vector $|\Omega\rangle$ can be defined by
\begin{equation}
    \mathcal{P}=\overline{\left\{\Delta^{1/4} a  \Omega\,;\qquad a \in A_+ \right\}}
\end{equation}
where $\Delta$ is the corresponding modular operator. It satisfies the following properties:
\begin{prop}\label{prop:cone}
Let $A$ be a von Neumann algebra and $\Omega$ a cyclic and separating vector. We have
    \begin{enumerate}
        \item $ \mathcal{P}=\overline{\left\{a J a J \Omega\,;\qquad a \in \mathcal{A} \right\}}$
        \item $\Delta^{it}\mathcal{P}=\mathcal{P}$, $\qquad t\in \mathbb{R}$
        \item $J \Psi = \Psi$, $\qquad \Psi \in \mathcal{P}$
        \item   $a J a J \mathcal{P}\subset \mathcal{P}$
        \item $\mathcal{P}$ is a pointed self-dual cone with span$(\mathcal{P})=\mathcal{H}$
        \item If $\Psi \in \mathcal{P}$ then $\Psi$ is cyclic for $A$ if and only if it is separating for $A$
        \item If $\Psi  \in \mathcal{P}$ is cyclic for $A$ (and therefore separating), then the modular conjugation and natural cone associated with $(A,\Psi)$ coincides with $J$ and $\mathcal{P}$.
        \item For every normal state $\phi$ on $A$ the exists  a unique vector $\Psi_\phi \in \mathcal{P}$  such that $\phi(a)=\langle \Psi_\phi|a|\Psi_\phi\rangle$ for all $a\in A$. 
    \end{enumerate}
\end{prop}

\bibliographystyle{JHEP}
\bibliography{bibliography}
\end{document}